\documentclass[12pt,letterpaper]{article}

\usepackage{dsfont}
\usepackage{graphicx}

\usepackage{amsfonts}
\usepackage{amssymb}
\usepackage{amsmath}
\usepackage{appendix}

\def \ov {\over}

\newcommand{\be}{\begin{equation}}
\newcommand{\ee}{\end{equation}}
\newcommand{\eel}[1]{\label{#1}\end{equation}}
\newcommand{\bea}{\begin{eqnarray}}
\newcommand{\eea}{\end{eqnarray}}
\newcommand{\eeal}[1]{\label{#1}\end{eqnarray}}

\renewcommand{\thefootnote}{\fnsymbol{footnote}}
\numberwithin{equation}{section}

\def\appendix#1{
  \addtocounter{section}{-3}
  \setcounter{equation}{1}
  \renewcommand{\thesection}{\Alph{section}}
  \section*{Appendix \thesection\protect\indent \parbox[t]{11.15cm}
  {#1} }
  \addcontentsline{toc}{section}{Appendix \thesection\ \ \ #1}
  }
\def \ov {\over}

\def\be{\begin{equation}}
\def\ee{\end{equation}}

\date{}
\begin{document}
%\draft
\begin{titlepage}
\hfill MCTP-11-37\\

\begin{center}
%\vskip 2.5 cm \vskip 1 cm
{\Large \bf On Field Theory Thermalization}
\vskip .7 cm
{\Large \bf from  Gravitational Collapse}
\end{center}

\vskip .7 cm

\vskip 1 cm
\begin{center}
{ \large David Garfinkle${}^{1,2}$, Leopoldo A. Pando Zayas${}^2$ and Dori Reichmann${}^2$}
\end{center}
\vskip .4cm \centerline{\it ${}^1$  Department of Physics }
\centerline{ \it Oakland University}
\centerline{\it Rochester, MI 48309}

\vskip .4cm \centerline{\it ${}^2$ Michigan Center for Theoretical
Physics}
\centerline{ \it Randall Laboratory of Physics, The University of
Michigan}
\centerline{\it Ann Arbor, MI 48109-1120}

\vskip 1 cm

\vskip 1.5 cm
\begin{abstract}
Motivated by its field theory interpretation, we study gravitational collapse of a minimally coupled
massless scalar field in Einstein gravity with a negative cosmological constant. After demonstrating the
accuracy of the numerical algorithm for the questions we are interested in, we investigate various aspects
of the apparent horizon formation. In particular, we study the time and radius of the apparent horizon formed
as functions of the initial Gaussian profile for the scalar field. We comment on several aspects of the dual field theory picture.
\end{abstract}

\end{titlepage}

\setcounter{page}{1} \renewcommand{\thefootnote}{\arabic{footnote}}
\setcounter{footnote}{0}
\def \N{{\cal N}}
\def \ov {\over}

%%%%%%%%%%%%%%%%%%%%%%%%%%%%%%%%%%%%%%%%%%%%%%%%%%%%%%%%%%%%%%%
\section{ Introduction}
%%%%%%%%%%%%%%%%%%%%%%%%%%%%%%%%%%%%%%%%%%%%%%%%%%%%%%%%%%%%%%%

Understanding cooperative phenomena far from equilibrium poses one of the most challenging problems of present day physics.  One  source of
methods is  provided by the AdS/CFT correspondence  \cite{Maldacena:1997re,Witten:1998qj,Gubser:1998bc,Aharony:1999ti} which  identifies
a field theory without gravity with a string theory. In the low energy limit of string theory one can use the appropriate
generalization of Einstein's equations to follow the evolution in time of the fields in the dual field theory. This is particularly suitable for the description of strongly coupled field theories near a conformal fixed point. In this context the
gauge/gravity duality opens a particularly important window: to study far from
equilibrium phenomena one needs to study the evolution of Einstein's equation with appropriate boundary conditions.

The AdS/CFT correspondence was readily expanded to include field theories at finite temperature \cite{Witten:1998zw}. In this
context a field theory in equilibrium at finite temperature is dual to a black hole in asymptotically $AdS$ spacetime. One very
important development has been the establishment of the correspondence at the level of small fluctuations. More precisely, the
application of linear response theory for the near equilibrium region, {\it id est},  long wave lengths  and low energies with local fluid
variables varying very slowly compared to microscopic scales (see \cite{Son:2007vk,Hubeny:2010ry,Hubeny:2011hd} for reviews and
complete list of references). The next frontier comes from the expectation that the  evolution of spacetimes with the formation of black
hole horizons is equivalent to non-equilibrium dynamics and evolution towards thermalization in field theory.

Besides the general reasons to study far from equilibrium phenomena using the gauge/gravity correspondence one practical motivation
comes from the RHIC  and LHC experiments. Two crucial points are worth highlighting: (i) the observed Quark-Gluon Plasma (QGP) is
strongly coupled and (ii) hydrodynamical description  fits in a wide range of processes.  Theoretical and experimental developments
indicate that the QGP produced at RHIC is a strongly interacting liquid rather than the weakly interacting gas of quarks and
gluons that was previously expected  \cite{Shuryak:2003xe,Shuryak:2004cy,Heinz:2004pj}. A crucial piece of experimental evidence
pointing to the need for methods applicable to strongly coupled theories comes from the fact that the produced plasma locally
isotropizes over a time scale of $\tau_{iso}\le 1 fm/c \,\,$( where $c$ is the speed of light). The dynamics of such rapid isotropization in a far-from-equilibrium
non-Abelian plasma can not be described with the standard methods of field theory or hydrodynamics \cite{Muller:2008zzm}. In
this manuscript we propose to study such rapid thermalization via its gravity dual -- gravitational collapse. In this publication
we largely expand, clarify and sharpen the results presented in our short letter \cite{Garfinkle:2011hm}.

String theory also provides a framework for
understanding some superconductors holographically \cite{Hartnoll:2008vx,Gubser:2008zu}
(see reviews \cite{Hartnoll:2009sz,Horowitz:2010gk}). Moreover, some hard to understand properties of condensed matter
system like the structure of non-Fermi liquids have been recently described using holographic
models \cite{Cubrovic:2009ye} \cite{Faulkner:2010zz} (see \cite{McGreevy:2009xe} for a review). The question of thermalization is also crucial in condensed matter setups. There is an active interest in the understanding of
the time evolution of a system following a quantum quench \cite{RigolNature,PhysRevLett.103.100403,Calabrese:2006rx,Cardy:2011zz}. A holographic
approach to this area is showing promise in lower dimensions \cite{AbajoArrastia:2010yt,Aparicio:2011zy}. The experimental
observation of rapid thermalization in two-species mixtures of quantum degenerate Bose and Fermi Gases \cite{PhysRevLett.88.160401}
constitutes another strong motivation for the study of thermalization using the gauge/gravity dictionary.  The holographic
description of these models always involves a black hole and the formation of such a black hole will help our
understanding of those systems at a more fundamental level.

In this paper  we present a full numerical analysis of the gravitational collapse of a massless minimally coupled scalar
field in the presence of a negative cosmological constant in five spacetime dimensions. We go beyond previous attempts
within the AdS/CFT approach that relied on perturbation theory \cite{Bhattacharyya:2009uu} or toy models for
quench  \cite{Balasubramanian:2010ce,Erdmenger:2011jb,Balasubramanian:2011ur}. Our focus is on properties of the
thermalization time.

An intuitive explanation for the rapid thermalization time at RHIC within the context of the gauge/gravity correspondence is as follows.
Local gauge-invariant operators are mostly sensitive to geometry near the boundary. As explained in \cite{Bhattacharyya:2009uu}, in
the large $N$ limit which is the region explored by the gauge/gravity correspondence, trace factorization ensures that the
expectation value of products equals the product of expectation values. Only one-point functions of gauge invariant
operators survive. Other studies supporting rapid thermalization in the context of the gauge/gravity correspondence
include \cite{Balasubramanian:2010ce,Balasubramanian:2011ur,Ebrahim:2010ra,Asplund:2011qj}.

The paper is organized as follows. In section \ref{sec:Setup} we set the stage for the problem, that is, we present the Ansatz and equations
of motion. Section \ref{sec:Numerical} contains details about the numerics involved and a brief discussion about the regime of
very small amplitudes and gravitational collapse after a few bounces of the scalar field. Section \ref{sec:bhformation} contains our main
new results from the gravity point of view where we present a systematic study of the time of formation of the apparent
horizon. Since our work could be of interests to various communities (string theory, numerical relativity and
non-equilibrium field theory), in section \ref{sec:Comparing} we survey the current literature for the benefit
of the reader and explain how our results extend previous ones in the literature. We discuss various aspects of
the field theory dual in section \ref{sec:ft}. We conclude in section \ref{sec:Conclusions}.  We also include
a series of appendices where we collected a few facts that will help interpreting the numerical results presented
in the main body of the paper; depending on the readers background they could be obvious.

%%%%%%%%%%%%%%%%%%%%%%%%%%%%%%%%%%%%%%%%%%%%%%%%%%%%%%%%%%%%%%%%%%%%%%%%%
\section{Collapse in asymptotically $AdS_{5}$ spaces}\label{sec:Setup}
We consider the minimal Einstein-scalar field action with a negative cosmological constant $\Lambda = -d(d-1)/(2L^2)$:
\be
\label{Eq:Action}
S=\int d^{d+1}x \sqrt{-g}\bigg[\frac{1}{2\kappa^2}\left(R+\frac{12}{L^2}\right) -\frac12 (\partial\phi)^2 -U(\phi)\bigg].
\ee
The Einstein and Klein-Gordon equations of motion are respectively:
\bea
G_{\mu\nu}-g_{\mu \nu}\frac{d(d-1)}{2L^2}&=&\kappa^2\bigg[\partial_\mu \phi \partial_\nu\phi-\frac12 g_{\mu\nu} (\partial\phi)^2 -g_{\mu\nu}U(\phi) \bigg] \label{eq:Einstein}  \\
\frac{1}{\sqrt{-g}}\partial_\nu \left(\sqrt{-g}g^{\mu\nu}\partial_\mu \phi\right)-\frac{\partial U(\phi)}{\partial \phi}&=&0. \label{eq:Klein-Gordon}
\eea
We tackle this system using numerical methods developed and tested in \cite{Garfinkle:2004pw}, \cite{Garfinkle:2004sx} which we will
describe in more detail below. We choose the metric Ansatz to be
\be
\label{eq:metric}
ds^2 = - \alpha^2 dt^2+ a^2 dr^2 + r^2 (d \chi^2 +\sin^2 \chi (d\theta^2 +\sin^2\theta d\varphi^2 )),
\ee
where $\alpha$ and $a$ are each functions of only $t$ and $r$. For convenience, let us define
\be
X=\partial_r \phi, \qquad Y=\frac{a}{\alpha}\partial_t \phi.
\ee
The Einstein equations can be conveniently written as:
\bea
\partial_r a =\frac{a}{r}(1-a^2)&+&\frac{r\, a}{6}\bigg[X^2+Y^2+a^2(2U-\frac{12}{L^2})\bigg], \label{eq:a-prime} \\
\partial_r\ln (a\, \alpha)&=&\frac{r}{3}\left(X^2+Y^2\right), \label{eq:a-alpha-prime} \\
\frac{3}{r}\partial_t a&=&\alpha\, X\, Y. \label{eq:dot-a}
\eea
Here we are using units where $\kappa =1$. The Klein-Gordon equation  takes the form
\be
\label{eq:Klein-Gordon}
\partial_t Y=\frac{1}{r^3}\partial_r\left(r^3\frac{\alpha}{a}X\right)-\alpha\, a\, \partial_{\phi}U.
\ee
 The main difference of the system given by equations (\ref{eq:a-prime}, \ref{eq:a-alpha-prime} and \ref{eq:Klein-Gordon}) with respect to
 the treatment presented in \cite{Garfinkle:2004pw} and \cite{Garfinkle:2004sx} lies in the powers of $r$ that appear in $AdS_4$ there as opposed to $AdS_5$ here.  For example, in
 equation (\ref{eq:Klein-Gordon}) we have a cubic power of $r$ rather than a square one. This higher power of $r$ leads to a more
 singular behavior near $r=0$ which presents a numerical challenge.

It is worth pointing out that  equation (\ref{eq:dot-a}) can be shown to be  automatically satisfied when the other equations are
satisfied.  Numerically, that equation is used as a check to verify that the code is working properly. Namely the equation should
be satisfied up to a numerical error due to the finite grid spacing. In the next section we will focus on this constraint in a fair
amount of details.

%%%%%%%%%%%%%%%%%%%%%%%%%%%%%%%%%%%%%%%%%%%%%%%%%%%%
\section{Comments on numerics}\label{sec:Numerical}
%%%%%%%%%%%%%%%%%%%%%%%%%%%%%%%%%%%%%%%%%%%%%%%%%%%%%%%%%%%
We consider a massless scalar field, {\it i.e.}, $U=0$. We choose initial data of the form $\phi = A \exp (- {{(r- {r_0})}^2}/ \sigma^2 )$ where the amplitude $A$, center of the scalar
field profile $r_0$ and profile width $\sigma$ are constants.  This determines $X$ through $X={\partial _r}\phi$ and we choose
$Y=X$ initially so that the wave starts out purely ingoing.
We choose $L=1$ and for the spatial grid to be $0 \le r \le {r_{\rm max}}$ where $r_{\rm max}$ is a constant.  For the simulations
done in this work we use ${r_{\rm max}}=10$.  Since in anti-deSitter spacetime, more resolution is needed at small $r$ we do not
choose the spatial points to be evenly spaced.  Rather, for a simulation with $N$ spatial points, the value of $r$ at grid point
$i$ is ${r_i} = \sinh (k (i-1))$ where $k$ is a constant chosen so that ${r_N} = {r_{\rm max}}$. In the graphs below we use $N$ as a measure of the resolution of the simulation.

For sufficiently large amplitudes the collapse process results in the formation of a black hole.  This is signalled by the
formation of a marginally outer trapped surface (also called an apparent horizon),
where the outgoing null geodesics cease to diverge from each other and instead
begin to converge.  In spherically symmetric spacetimes, an apparent horizon has the property that
$\nabla^a r \nabla_a r=0$.  In our coordinate system this would mean $a^{-2}\to 0$.  The attentive reader whose intuition comes mostly from analytical solutions
would expect that the horizon could as well be defined as $g_{tt}\to 0$ which means  $\alpha \to 0$ near the horizon. That is not so in our situation as we stop the simulation based solely on the considerations of an apparent horizon discussed above and those considerations are
independent of $\alpha$.

More precisely, our coordinate system
breaks down when an apparent horizon forms, and so our signal that a black hole is forming is that
$a \to \infty$ in the simulations.  Actually, it suffices to set a fairly moderate value for the maximum allowed value of $a$,
which we denote $a_{\rm max}$ and to stop the simulation whenever $a$ reaches $a_{\rm max}$, noting that a black hole has formed
at that time.  The position $r_{AH}$ at which $a={a_{\rm max}}$ gives us the size of the black hole and allows us to estimate its mass $M_{AH}=3\pi^2 (r_{AH}^2 +r_{AH}^4/L^2)$. Another notion of mass arises from the energy momentum tensor. This is the notion that best fits with the ADM mass, it is conserved, we denote it as asymptotic mass or total mass and can be computed as follows

\be
\label{eq:M_BH}
M_{total}=\pi^2\int\limits_0^{r_{max}}\frac{r^3}{a^2}\left(X^2+Y^2\right)dr.
\ee
There is no {\it a priori} reason to expect $M_{AH}$ and $M_{BH}$ to coincide as not all the matter density initially in the scalar participates in the formation of the apparent horizon.

%%%%%%%%%%%%%%%%%%%%%%%%%%%%%%%%%%%%%%%%%%%%%%%%%%%%%%%%%%%%%%%%%%%%%%%%%%%%%%%%%%%%%%%%
\subsection{Understanding the numerical accuracy}\label{sec:accurary}

The code used in the simulations was originally used in a 4-d context in \cite{Garfinkle:2004pw}, \cite{Garfinkle:2004sx}. The precise version
used here produced the results presented in \cite{Garfinkle:2011hm}.

\begin{figure}[htp]
\begin{center}
\includegraphics[width=4.0in]{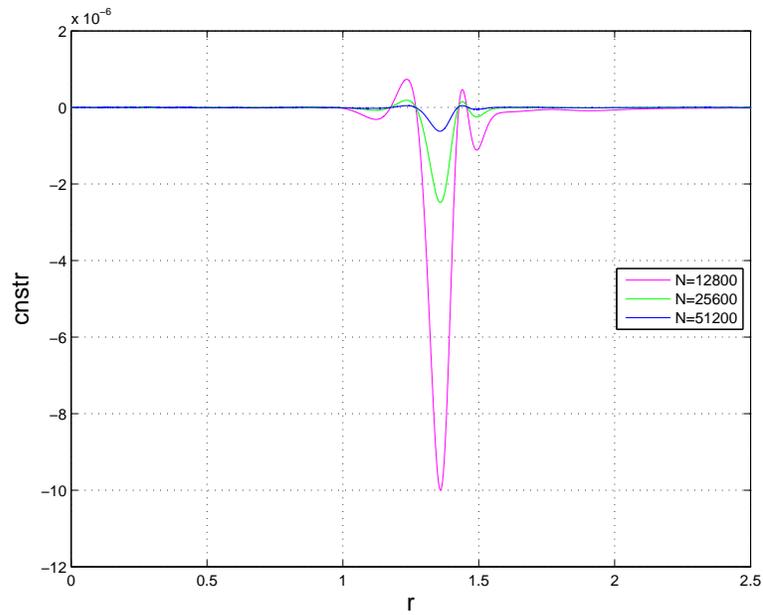}
\caption{\label{cnstrfig} The constraint quantity at various grid sizes. }
\end{center}
\end{figure}

In our previous work \cite{Garfinkle:2011hm}, we tested that the code is working properly by considering the constraint quantity
${\cal C} \equiv {\partial _t} a - r \alpha X Y /3$.  Note that it follows from eqn. (\ref{eq:dot-a}) that $\cal C$ vanishes.
However, the finite size of the grid spacing means that $\cal C$ will not vanish exactly in the simulation, but that instead in
a properly working simulation, a
smaller grid spacing should lead to a smaller $\cal C$.  In fig. (\ref{cnstrfig}) the result of a test of this sort is shown.
We have plotted $\cal C$ for simulations with 12800, 25600 and 51200 grid points.  In all cases the simulations have $A=0.02, \, {r_0}=4.0$ and $\sigma = 1.5$ and the simulations are run to a time of $t=0.454$. A gross estimate of the error directly from the figure suggest that error decreases by a factor of $(1/2)^2$ pointing to a second order convergent code.

\begin{figure}
\begin{center}
\includegraphics[width=4.0in]{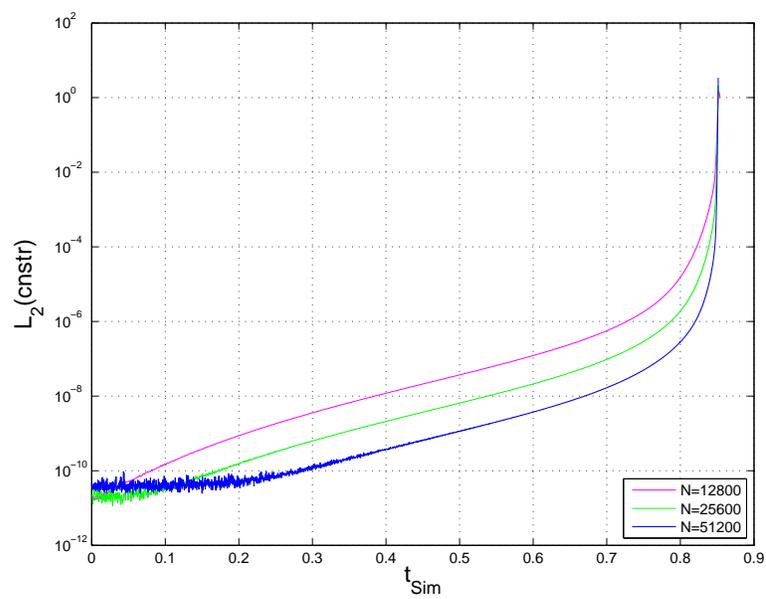}
\caption{\label{L2Norms} The $L_2$ norm of the constraint for various values of the resolution.}
\end{center}
\end{figure}

In this paper we consider simulations with  12800, 25600 and 51200 grid points respectively. As a test of the accuracy we use a
more invariant one. Namely, in fig. (\ref{L2Norms}) we plot the $L_2$ norm of the constraint as a
function of time. The key feature that we want to point out is the value of the $L_2$ norm of the constraint
as a function of the grid resolution. The value itself is not the most important aspect but the way that
it decreases as the number of points in the grid increases.

There are a few observations that can be made directly from fig. (\ref{L2Norms}). First, notice that we lose accuracy right at the time of
formation of the apparent horizon corresponding to the almost vertical asymptote. What follows from the figure is that the simulation is
very stable in determining the value of the apparent horizon as they all shoot up at the same value of time. A disastrous situation would
have corresponded to losing accuracy at times that are significantly different from the
apparent horizon formation time. It is also gratifying to see that the higher the resolution the closer we get to
the time of the apparent horizon formation maintaining higher accuracy. We could, in principle, estimate the error in quoting values for times
in our simulation based on the
behavior of the $L_2$ norms described in the figure.

\begin{figure}
\begin{center}
\includegraphics[width=4.0in]{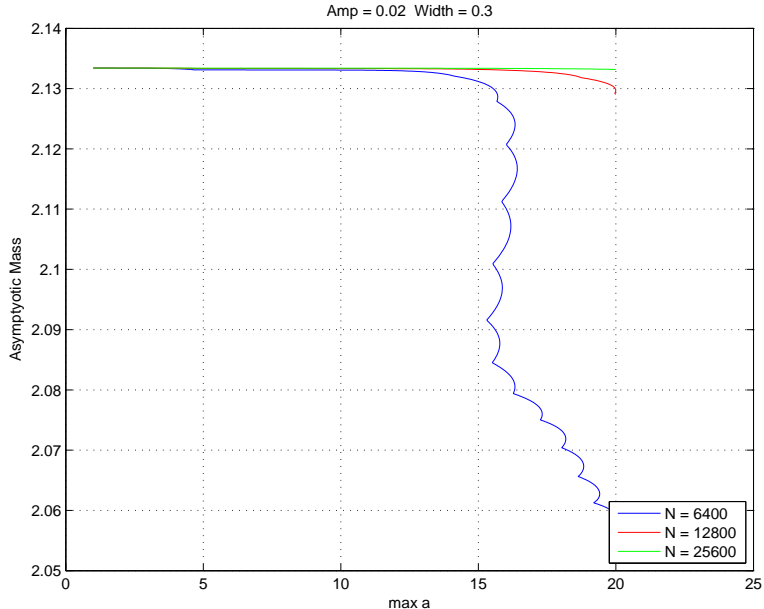}
\caption{\label{Mass-amax} Dependence of the asymptotic mass (see equation \ref{asymptotic_mass})  as a function of the maximum value of $a(r,t)$ which is used to determined apparent horizon formation. The asymptotic mass is expected to be constant and thus the plot can be understand as a test for the accuracy of the code. }
\end{center}
\end{figure}

We have also performed more dynamical tests on the code. For example, we have checked stability with respect to the definition of the
horizon $a_{max}$ by increasing its value tenfold and checking that the resulting radius and time of the formation of the apparent horizon
are not affected to the precision we are working. In figure (\ref{Mass-amax}) we plot the value of the asymptotic mass for various values of the resolution as a function of $a_{max}$. The precise definition of the asymptotic mass is given in equation (\ref{asymptotic_mass}); this quantity is constant in the simulation. What figure (\ref{Mass-amax}) shows is that for small values of $a_{max}$ the value of the mass is indeed constant as we increase $a_{max}$ we  see that each simulation breaks but the point at which they break can postpone by increasing the resolution. In a more practical sense we are are following the maximum value of the function $a(r,t)$, that is the plot (\ref{Mass-amax}) can be also interpreted as related to the simulation time at which, due to the raising value of $a(r,t)$ we lose accuracy. We also checked that the time and position of the apparent horizon formation were  stable to
the increment in resolution.

%%%%%%%%%%%%%%%%%%%%%%%%%%%%%%%%%%%%%%%%%%%%%%%%%%%%%%%%%%%%%%%%%%%%%%%%%%%%%%%%%%%%%%%%
\subsection{Oscillatory motion of the scalar field}

\begin{figure}[htp]
\begin{center}
\includegraphics[width=4.0in]{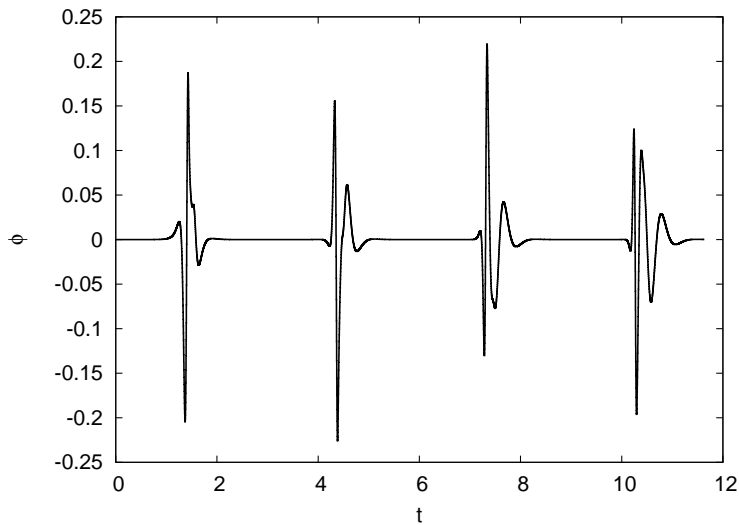}
\caption{\label{phi0fig} $\phi(0,t)$ for a small amplitude simulation. }
\end{center}
\end{figure}

For collapse in asymptotically flat spacetimes, small amplitude initial data leads to a wave that is initially ingoing, then
undergoes interference near the center and becomes an outgoing wave.  For collapse in asymptotically $AdS$ spacetimes,
we would expect similar behavior, except that the outgoing wave should then reach the $AdS$ boundary, bounce and become
ingoing again, leading to another bounce, and so on.  And indeed, this is what we find.  Fig. (\ref{phi0fig}) shows the value
of the scalar field at the center as a function of time.  ($\phi (0,t)$).  The parameters for this simulation are
$A=0.0002, \, {r_0} =4.0$ and $\sigma =1.5$.  This simulation was done with $6400$ grid points.  Note that there are particular
periods of time where the value of the scalar field at the center is non-negligible and that the middle of each such
time period is separated from the
next one by approximately $\pi$.  This is exactly what we would expect if the dynamics is mostly that of a massless scalar field
on a background anti-deSitter spacetime.  In the geometric optics limit, such a scalar field propagates along null geodesics, and
for anti-deSitter spacetime (with $L=1$) it takes a null geodesic a time of $\pi/2$ to propagate from the center to infinity (see appendix
\ref{app:geodesics} for details of geodesic calculations if needed.).

In our previous paper \cite{Garfinkle:2011hm} we concluded that this oscillatory behavior could go on indefinitely. A more detailed study of this
oscillatory behavior shows that the end outcome of this oscillatory behavior is black hole formation \cite{Bizon:2011gg}. In a follow up
note, Jalmuzna, Rostworowski and  Bizon \cite{Jalmuzna:2011qw} evolved our initial data in a more accurate setup
(allowing up to  $2^{17}+1$ grid points) and showed that a black hole forms after a time two orders of magnitude
greater than those considered originally in \cite{Garfinkle:2011hm}. An important proposal put forward  in \cite{Bizon:2011gg} states  that a
weak turbulent mechanism is responsible for the nonlinear instability of $AdS$ with respect to arbitrarily small perturbations. The weak turbulent
mechanism in question allows for a transfer of energy from larger to smaller scales. Corroboration for such mechanism was also  recently reported in
\cite{Dias:2011ss}. Notice that the incremental appearance of peaks in figure (\ref{phi0fig}) is a visual counterpart to the mechanism proposed in \cite{Bizon:2011gg}.

\begin{figure}[htp]
\begin{center}
\includegraphics[width=4.0in]{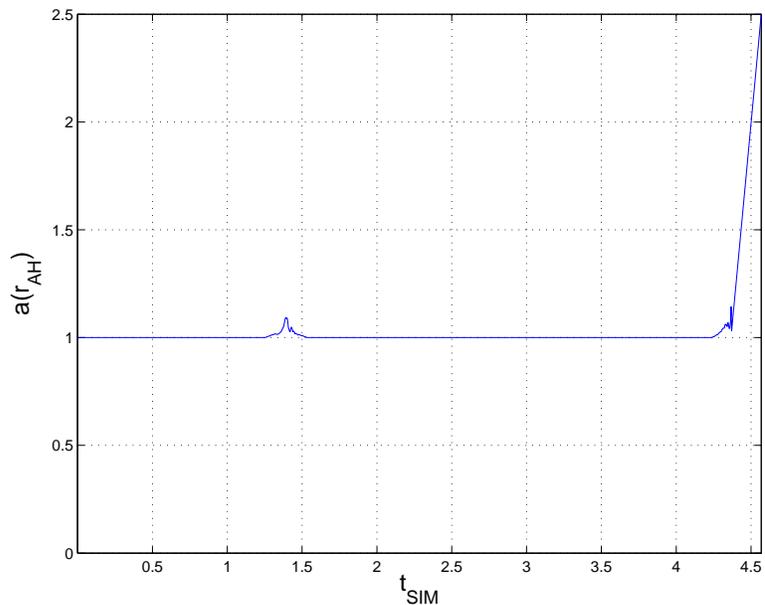}
\caption{\label{bounce-a} $a(r\approx r_{AH},t)$ for a small amplitude simulation that bounces once and then gravitationally collapses. }
\end{center}
\end{figure}

In this manuscript we limit ourselves to show that we have explicitly verified that, indeed, collapse is possible after a few bounces.  We show, in fig. \ref{bounce-a}, data evolved for the $A=0.00084, \sigma=1.5, r_0=4$ with a resolution of $N=12800$ points. We used \cite{Jalmuzna:2011qw} to find the most appropriate range of parameters for our purpose. Note that the scalar field bounces once and then gravitationally collapses, we show that in the figure by tracking the value of the metric function $a(r,t)$ near the position of eventual horizon formation. In the simulation a black hole forms at $r_{AH}=0.0049$ after a simulation time of $t_{SIM}=4.568$.

%%%%%%%%%%%%%%%%%%%%%%%%%%%%%%%%%%%%%%%%%%%%%%%%%%%%%%%%%%%%%%%%%%%%%%%%%%%%%%%%%%%%%%%%
\section{Black hole formation}\label{sec:bhformation}

The quantity that we are most interested in is how long the process of black hole formation takes, {\it i. e.},  we would like to
know how much time elapses
between the initial time and the time a marginally outer trapped surface forms.  For simulations it is natural to use the
coordinate $t$ as a measure of time and we have used it previously in \cite{Garfinkle:2011hm}. Here we continue to use that simplified notion of time, a more precise definition of the field theory time is described in appendix (\ref{app:times}) and amounts to correcting the coordinate time
to incorporate the right field theoretic normalization. We denote by
$t_{\rm AH}$ the simulation time $t$ at which an apparent horizon first forms.
Note that our coordinate time (and similarly the field theory time) $t$ has geometric meaning in that it is the coordinate that is orthogonal to the
area coordinate $r$. We discuss various relevant notions of time in more detail in appendix (\ref{app:times}).

\begin{figure}[ht]
\begin{minipage}[b]{0.5\linewidth}
\centering
\includegraphics[width=3.0in]{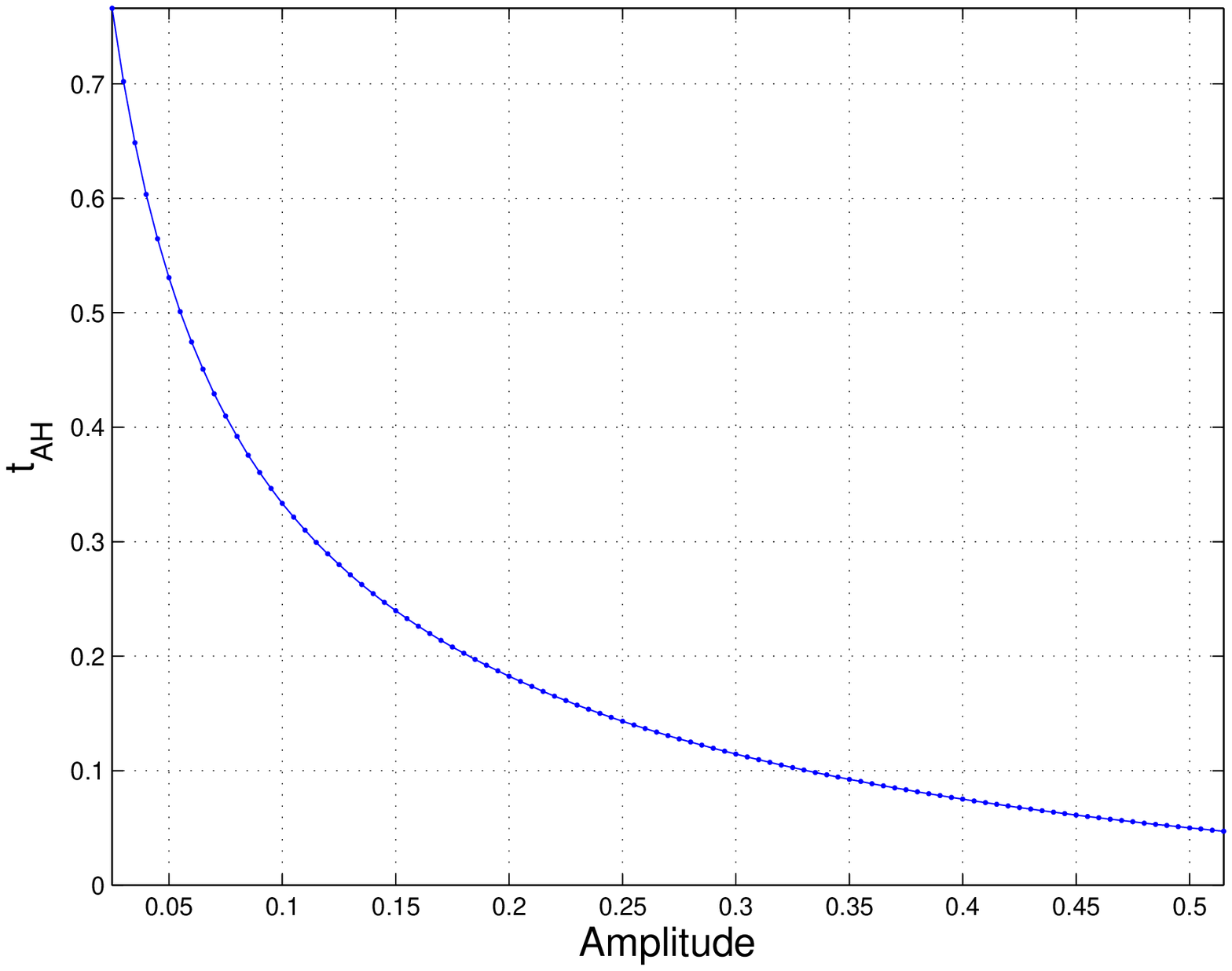}
\caption{\label{Atime} The dependence of $t_{\rm AH}$ on $A$ }
\end{minipage}
\hspace{0.5cm}
\begin{minipage}[b]{0.5\linewidth}
\centering
\includegraphics[width=3.0in]{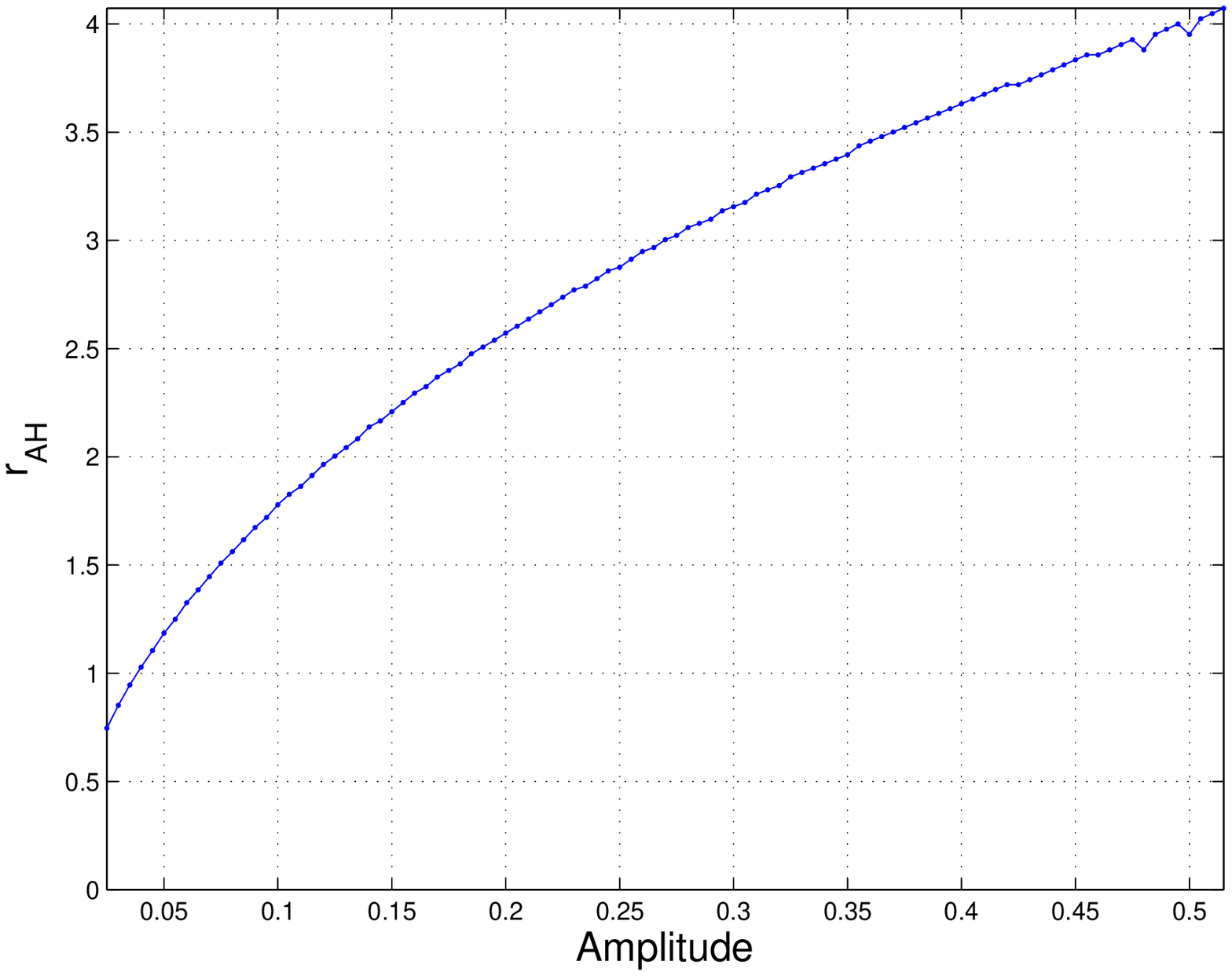}
\caption{\label{rVa-A} The dependence of $r_{\rm AH}$ on $A$ }
\end{minipage}
\end{figure}

We also want to know how the time of apparent horizon formation depends on the initial amplitude of the pulse $A$.
This dependence is shown in fig. (\ref{Atime}) which gives
$t_{\rm AH}$ for several simulations with different values of $A$.  For each of these simulations we have
$\sigma =1.5$ and ${r_0} =4.0$. These simulations were done with $51200$ grid points.   Similarly, in figure (\ref{rVa-A}) we plotted the dependence of the radius of the apparent horizon as a function of the amplitude. The masses of the
black holes formed in the simulations range from $23.5$ for the smallest amplitude to $8.52 \times 10^{3}$ for the largest amplitude, that is, about three orders of magnitude.

Some aspects of figures  (\ref{Atime}) and (\ref{rVa-A}) can be understood using the properties of null geodesics in
anti-deSitter spacetime. We are going to slowly build up our intuition on this account.

First, fig. (\ref{rVa-A}) shows that the radius of the apparent horizon grows with the amplitude $A$. This is intuitively clear as the higher the amplitude $A$ the higher the initial energy stored in the scalar field and consequently the higher the value of the asymptotic mass $M_{BH}$ defined
in equation (\ref{eq:M_BH}).

\begin{figure}[ht]
\begin{minipage}[b]{0.5\linewidth}
\centering
\includegraphics[width=3.0in]{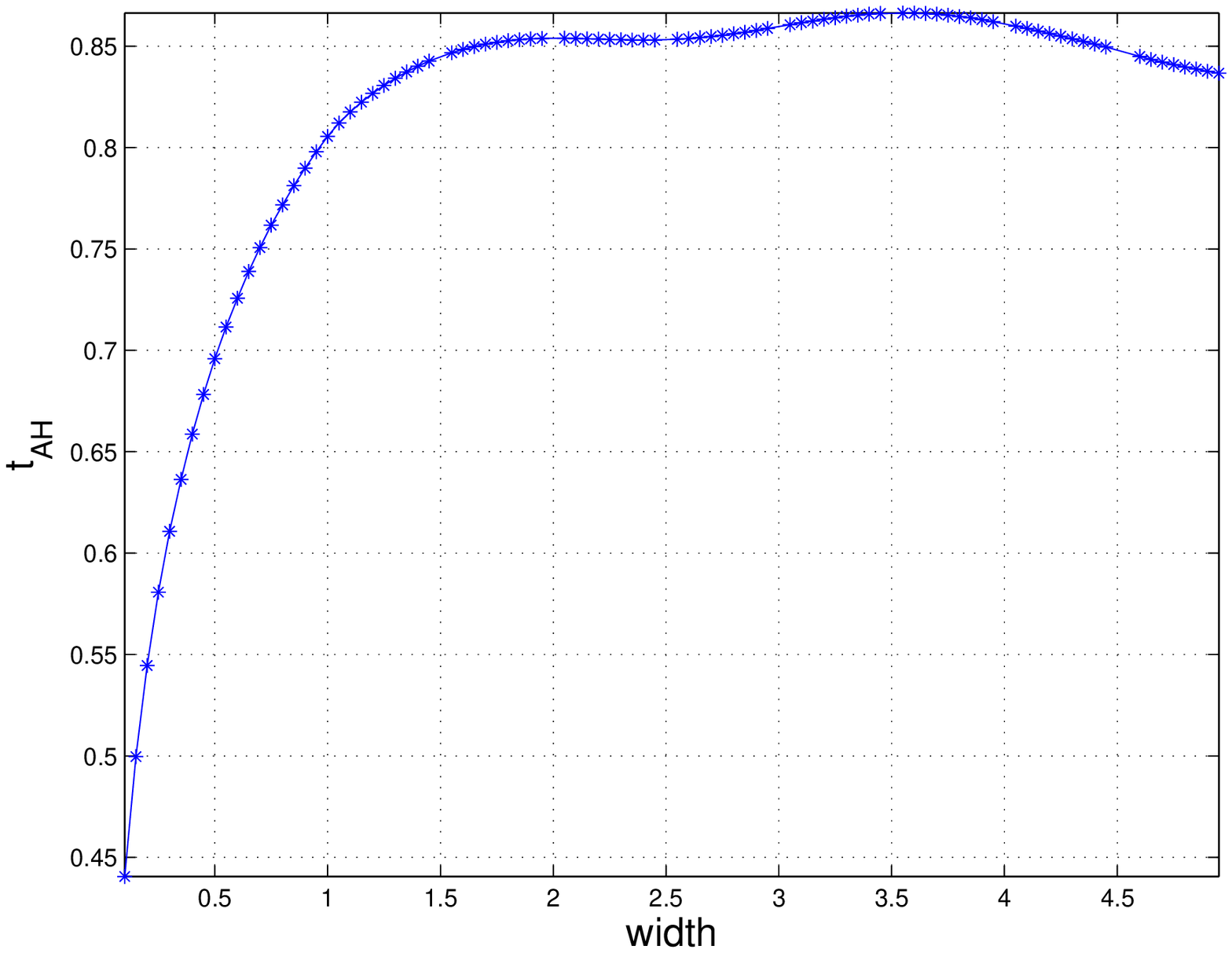}
\caption{\label{Wtime} The dependence of $t_{\rm AH}$ on $\sigma$ }
\end{minipage}
\hspace{0.5cm}
\begin{minipage}[b]{0.5\linewidth}
\centering
\includegraphics[width=3.0in]{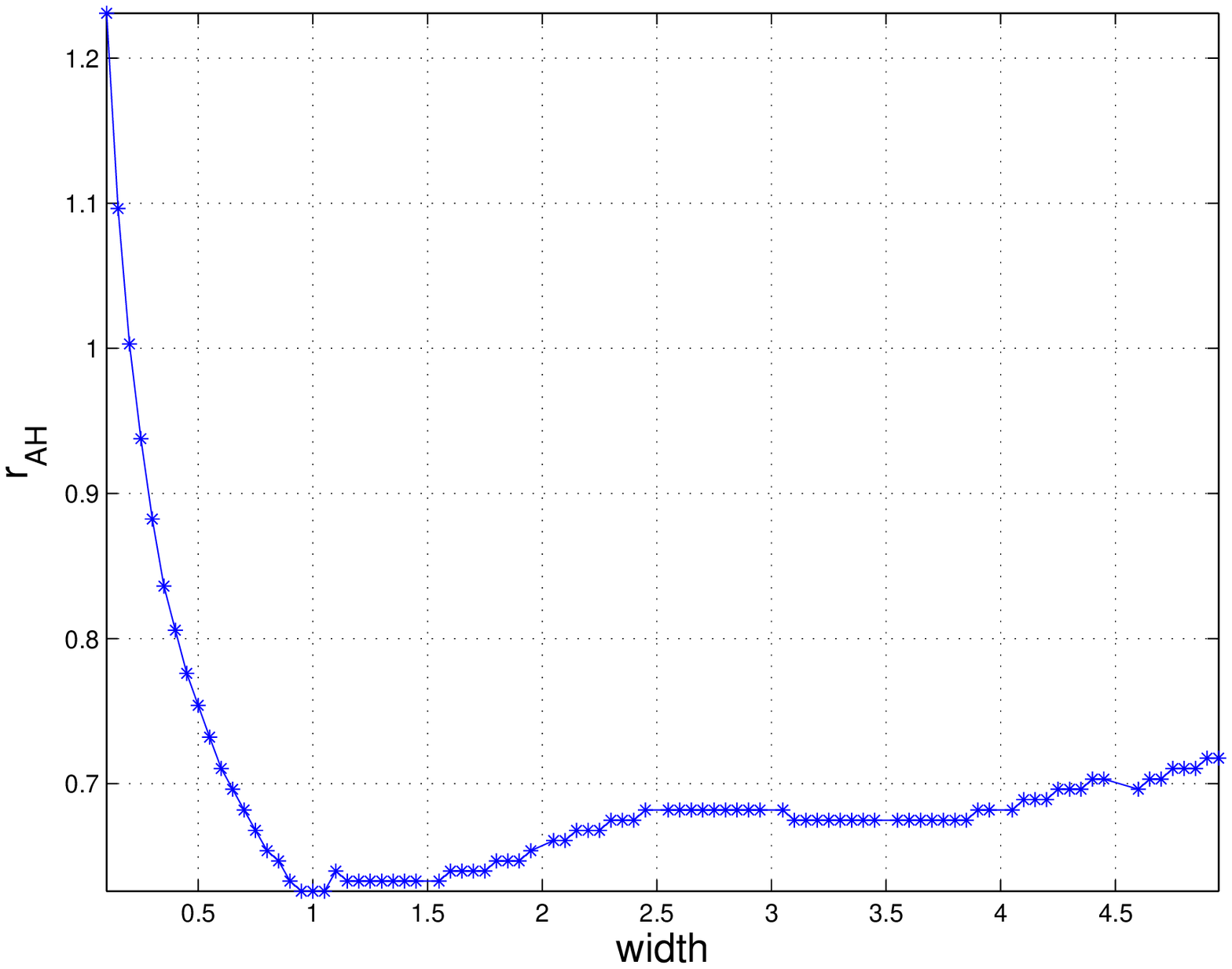}
\caption{\label{WrAH} The dependence of $r_{\rm AH}$ on $\sigma$.}
\end{minipage}
\end{figure}

To gain a better understanding of the time of black hole formation $t_{\rm AH}$ we also study its dependence on the initial width
of the pulse $\sigma$. This dependence is shown in fig. (\ref{Wtime}) which gives
$t_{\rm AH}$ for about a hundred simulations with different values of $\sigma$.  For each of these simulations we have
$A=0.02$ and $r_0 =4.0$.  These simulations were done with a radial resolution of $51200$ points.

The main intuition that follows from this graphs is that once the width has increased to a given value, the time of formation of the
apparent horizon remains approximately constant.  It is instructive to understand this result in a semi-analytic fashion. Rewriting the metric function $a$ as
\be
a^{-2}=1+\frac{r^2}{L^2}-\frac{ M(r,t)}{3\pi^2\, r^2},
\ee
we have that equation (\ref{eq:a-prime}) takes the following form for the local mass $M(r,t)$
\be
\label{asymptotic_mass}
\frac{\partial}{\partial r}M(r,t)=\pi^2\frac{r^3}{a^2}\left(X^2+Y^2\right).
\ee
The asymptotic mass called $M_{total}$ in equation (\ref{eq:M_BH}) is simply the limit as $r\to r_{max}$ of $M(r,t)$. We can estimate, in the thin-shell approximation, the dependence of $M$ on the width of the initial Gaussian profile.  Namely, in the thin-shell approximation ($\sigma \ll r_0$) we evaluate the above integral at $t=0$ where the $X=Y$ and we can approximate $a\sim 1/r$, then the leading term in $\sigma$ is
\be
M\approx 2 \pi^2\, A^2r_0^5\sqrt{\frac{\pi}{2}}\,\,\frac{1}{\sigma} +{\cal O}\left(\frac{\sigma}{r_0}\right)
\ee
This behavior of the mass leads to a dependence of the form $r_{AH}\sim \sigma^{-1/4}$ which is precisely what we see in figure (\ref{WrAH}) for small $\sigma$, in the range $\sigma \ge 1$  other contributions to $M$  flatten the curve.

This figure allows us to conclude that for a given amplitude $A$ increasing the width $\sigma$ in the original pulse amounts to decreasing the mass of the black hole that might eventually form.

The time it takes a null geodesic to travel from the center of the Gaussian profile at $r_0$ to the apparent horizon radius $r_{AH}$ can be explicitly calculated in $AdS$ and the result is (see appendix \ref{app:geodesics})

\be
\label{Eq:t_AH}
t_{AH}=L\left(\arctan\left(\frac{r_0}{L}\right)- \arctan\left(\frac{r_{AH}}{L}\right)\right).
\ee

\begin{figure}[htp]
\begin{center}
\includegraphics[width=5.0in]{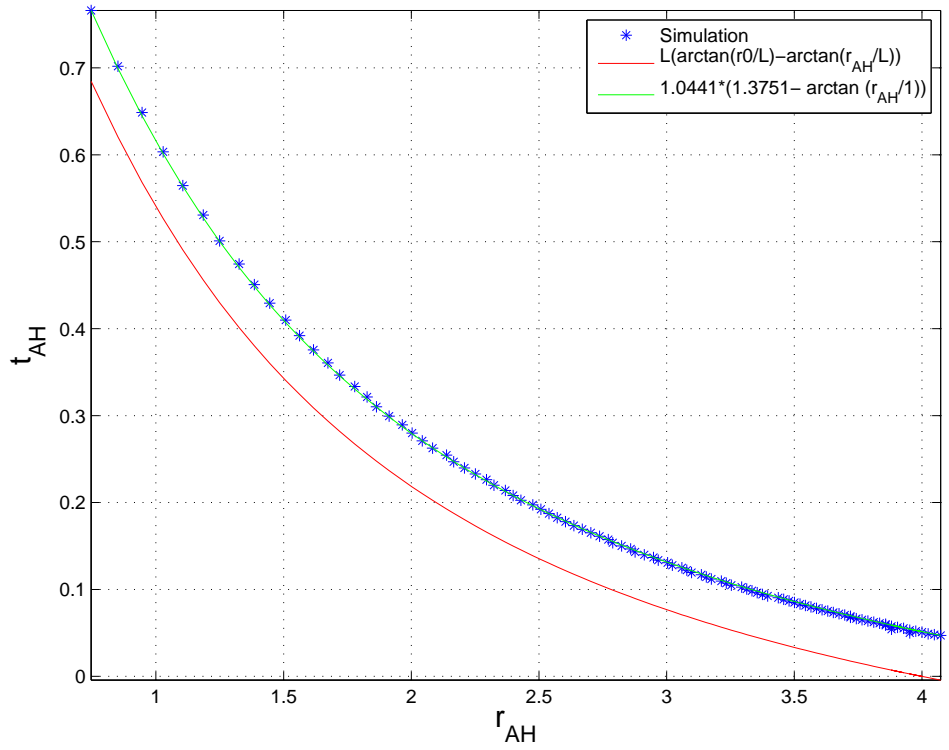}
\caption{\label{tAHVrAH} The dependence of $t_{\rm AH}$ on the radius of the apparent horizon }
\end{center}
\end{figure}

One of the main results of our paper is contained in the figure (\ref{tAHVrAH}). The figure contains the result of
about a hundred simulations, we have plotted $t_{AH}$ versus $r_{AH}$ from the simulations. In the same figure we
have superimpose the analytic expression for the time it takes a null geodesic to go from the center of the original
Gaussian packet $r_0$ to the horizon radius. We have also presented a best fit with an analytic form that closely
follows the result of the geometric approximation.

The intuition that the time of horizon formation would be closely determined by the null geodesic time was a prevalent
intuition in the literature. Our simulations show that the time of formation of an apparent horizon is still of the same order but slightly
bigger than the one given by a null geodesic as figure (\ref{tAHVrAH}) nicely shows. Clearly, the width in the original Gaussian is playing a prominent role. Since we
are solving the full nonlinear system of Einstein and Klein-Gordon equations we are able to take into consideration
configurations far from the thin-shell limit. The full understanding of the shape of the Gaussian profile in the
time of formation of the apparent horizon  will require more work. We will focus on the prevalent role of the width in what follows.

We finish this section by advancing an interpretation of the best fit to the data as presented in figure (\ref{tAHVrAH}). First, the value $1.3751$ turns out to be well approximated by $\arctan(5)$, this suggests that the {\it effective} center of the Gaussian profile closer this value than to $r_0=4$. This is a direct consequence of the properties of $AdS$. Second, we interpret the coefficient of $1.0441$ as the extra time that it takes a shell of finite width to collapse. Thus figure (\ref{tAHVrAH}) with its best fit curve can be understood as summarizing the effects of a null geodesic falling from an effective distance of $r_0=5$ in AdS and also the effect of the extra time that takes the Gaussian to shrink.

%%%%%%%%%%%%%%%%%%%%%%%%%%%%%%%%%%%%%%%%%%%%%%%%%%%%%%%%%%%%%%%%%%%%%%%%%%%%%%%%%%%%%%%%%%%%%%%%%%%%%%%%%%%%%%%%%%%%%%%%%%
\subsection{A quantitative picture of gravitational collapse}

\begin{figure}[htp]
\begin{center}
\includegraphics[width=4.0in]{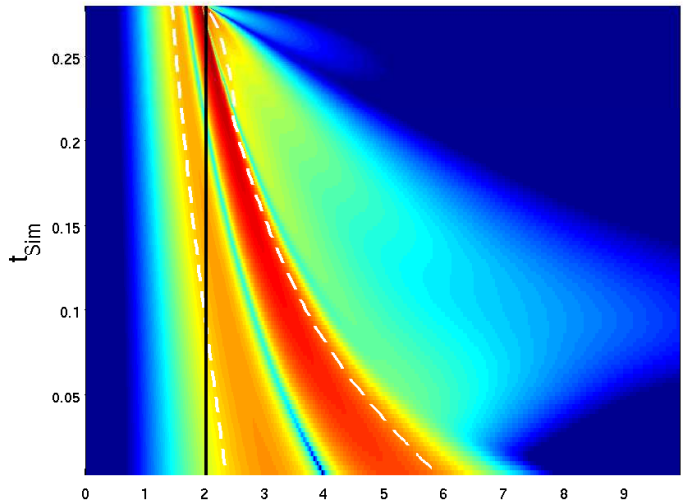}
\caption{\label{shell-norm-21} Time evolution and radial distribution of the energy density stored in the scalar field with Gaussian profile as the apparent horizon forms.}
\end{center}
\end{figure}

With the hope of clarifying the main properties of the space-time during the gravitational collapse we present fig. (\ref{shell-norm-21}).
In the figure we plot the Klein-Gordon energy density.  Consider a unit timelike vector $n=\alpha^{-1}\partial/\partial t$, then  the energy density is defined as (see appendix \ref{app:EoM})
\bea
\rho &=& T_{\mu\nu}n^\mu n^\nu \nonumber \\
&=& \alpha^{-2}\left(\partial_t \phi \partial_t \phi -\alpha^2 \frac{6}{L^2} +\frac{\alpha^2}{2}\partial^\nu \phi \partial_\nu \phi\right) \nonumber \\
&=&-\frac{6}{L^2} +\frac{1}{2 a^2}\left(X^2 +Y^2\right)
\eea
Based on this energy density we will plot a slightly different quantity which focuses on the scalar energy
\be
\rho_{Norm}(r,t)=\frac{\frac{1}{2 a^2}\left(X^2 +Y^2\right)}{\int\limits_0^{r_{max}}\frac{dr}{2 a^2}\left(X^2 +Y^2\right)}.
\ee
The advantage of this quantity is that for each value of time we normalize the energy to its total value on that line. This is simply a choice of presentation, otherwise, it becomes hard to follow what regions are really dense at a given time.

The simulation presented in fig. (\ref{shell-norm-21})  was conducted for $A=0.125, \sigma=1.5, r_0=4$ and with $N=51200$ grid points.
In fig. (\ref{shell-norm-21}) the vertical axis represents time and the horizontal axis represents position $r$. There are various properties to note about our measure of energy density for the scalar field. The energy density is proportional to the derivative of the scalar field and, from the initial condition, derivative in time and in
the radial directions are correlated, therefore,
we see a dark blue spot at $r=r_0=4$ corresponding to a zero of the energy density at $t=0$. One might expect a symmetry around that point, however the expression for energy contains an effective extra power of $r^2$ and therefore shifts
the maximum value to $r=5$. The black vertical line in figure (\ref{shell-norm-21}) represents the spatial position where the apparent horizon eventually forms. The horizon only actually forms at a later time in the simulation but we draw this spatial black vertical line to  highlight the position. To guide the reader we have also drawn two white dashed lines between which 90\% of the Klein-Gordon energy density of the scalar field is contained.

What we see clearly
is that, to a good approximation, the front  part of the scalar wave
crosses the apparent horizon and the second
part moves effectively in a Schwarzschild-AdS-like  background and experiences a marked stretching.

Let us also remark in this section a fact that will be important in the field theory interpretation of our results. The deep blue regions
in figure
(\ref{shell-norm-21}) are regions where the energy density of the scalar field is basically zero. The equations of motion, or more
generally Birkhoff's theorem, imply that the spacetimes in those regions can only be $AdS_5$ or the Schwarzschild black hole in $AdS_5$.
Therefore a boundary configuration that explores only the asymptotic region will only detect $AdS_5$ or the Schwarzschild black hole in $AdS_5$.

\begin{figure}[htp]
\begin{center}
\includegraphics[width=4.0in]{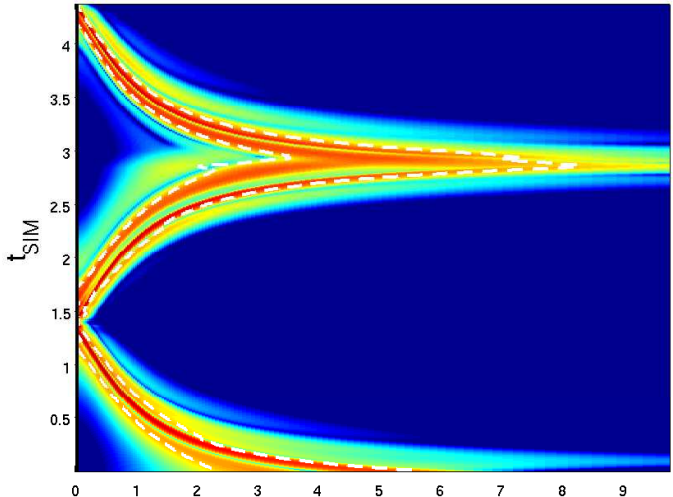}
\caption{\label{shell-norm-bounce} Time evolution and radial distribution of the energy density stored in the scalar field with Gaussian profile as the apparent horizon forms at a very small radius $r_{AH}=0.0049$ after a bounce of the scalar field.}
\end{center}
\end{figure}

For completeness, in fig. (\ref{shell-norm-bounce}), we also present the energy density graph for a situation where the scalar field bounces once and then collapses. The parameters of the simulation are exactly those presented in the previous section. Namely, $A=0.00084, \sigma=1.5, r_0=4$ with a resolution of $N=12800$ points. It is worth noticing that we have performed the same calculation with other resolutions and the results are indistinguishable from those presented here.

%%%%%%%%%%%%%%%%%%%%%%%%%%%%%%%%%%%%%%%%%%%%%%%%%%%%%%%%%%%%%%%%%%%%%%%%%%%%%%%%%%%%%%%%%%%%%%%%%%%%%%%%%%%%%%%%%%%%%%%%%%
\subsection{Comparison with previous results in the literature}\label{sec:Comparing}
The question of gravitational collapse in asymptotically AdS spaces has attracted a lot of attention largely due to its application
in the AdS/CFT correspondence. Some of the first works, for example  \cite{Danielsson:1999zt,Danielsson:1999fa,Giddings:1999zu} discussed
the setup and its implication in the context of the gauge/gravity  correspondence. An interesting model for collapsing  matter shells
in analogy with the Oppenheimer-Snyder solution was discussed in \cite{Giddings:2001ii}. Other versions of the Oppenheimer-Snyder
model in asymptotically AdS spaces were also considered in \cite{Alberghi:2003ce,Alberghi:2003pr}.   In the numerical relativity
community, similar problems have received a lot of attention. For example, collapse in asymptotically $AdS_3$ has been addressed
with beautiful results in
\cite{Birmingham:1999yt,Husain:2000vm,Pretorius:2000yu,Birmingham:2001hc}. Similarly, some work in  $AdS_4$ was
presented in, for example,  \cite{Husain:2002nk}. More general dimensions have been briefly considered
in \cite{Birukou:2002ge}. Aspects of potential cosmic censorship violations in asymptotically $AdS_4$
collapse have also been addressed by Garfinkle \cite{Garfinkle:2004pw,Garfinkle:2004sx}. Recently, a
discussion of black holes in a box, which is an important step in the full construction due to the fact
that $AdS$ is not globally hyperbolic, was presented in \cite{Witek:2010zz,Witek:2010qc}. More recently the question of turbulent
instability in asymptotically AdS spaces has been studied in the spherically symmetric \cite{Bizon:2011gg,Jalmuzna:2011qw} and
non-spherically symmetric \cite{Dias:2011ss} setups; the complete set of implications for the AdS/CFT remains a truly interesting problem.
There are other approaches that have been pursued with similar field theoretic intent. For example, a different
way of setting up the problem is due to Chesler and Yaffe \cite{Chesler:2008hg,Chesler:2009cy} who considered
a time dependent boundary metric $g_{\mu\nu}^B$ which creates gravitational radiation that propagates from
the boundary to the bulk. This gravitational radiation leads to the formation of a horizon. The anisotropic nature of the
problem is particularly relevant for applications to collision situations  but also
for the quantum quench setup discussed recently directly in field theory \cite{RigolNature},
\cite{PhysRevLett.103.100403}, \cite{Sotiriadis:2010si}, \cite{Calabrese:2007rg}, \cite{Calabrese:2006rx},  \cite{Cardy:2011zz}. Another
example of a different approach has been presented in a  beautiful recent paper \cite{Heller:2011ju} where the authors studied
the approach to the transition to the hydrodynamic regime using newly developed techniques for higher order hydrodynamics
in the context of the boost-invariant plasma originally proposed in \cite{Janik:2005zt}. Namely, the authors
of \cite{Heller:2011ju} studied the far from equilibrium dynamics of a system with boost-invariant symmetry, arguably
the dual of Bjorken flow
\cite{Bjorken:1982qr} in field theory; they showed that using higher order hydrodynamics the approach to the
linear hydrodynamic regime occurs in time scales inversely proportional to the equilibrium temperature.

%%%%%%%%%%%%%%%%%%%%%%%%%%%%%%%%%%%%%%%%%%%%%%%%%%%%%%%%%%%%%%%%%%%%%%%%%%%%%%%%%%%%%%%%%%%%%%%%%%%%%%%%%%%%%%%%%%%
{\it \large The thin Vaidya shell approximation -- }
Of the many setups discussed in the literature, the one that can be most directly compared to our work here is the one
presented in \cite{Bhattacharyya:2009uu}. This work provided, to our knowledge, the first systematic analysis of the
thin shell Vaidya collapse as the dual of rapid thermalization in field theory.

Let us briefly discuss the setup in \cite{Bhattacharyya:2009uu} which  considered an action of the form (\ref{Eq:Action}).
Their metric is spherically symmetric and is given in Vaidya type coordinates
\bea
ds^2&=&2dr dv -g(r,v)dv^2 +f^2(r,v)d {\Omega ^2}
\nonumber \\
\phi &=& \phi(r,v).
\eea
where $d {\Omega ^2}$ is the line element of the unit $S^3$. The crucial physical information is stored in the
scalar field profile which is $\phi_0(v) <\epsilon$ for $0<v<T$ and otherwise vanishes.
In \cite{Bhattacharyya:2009uu}, a perturbation theory in $\epsilon$ was developed.  The situation we treat is not precisely
the same as that of \cite{Bhattacharyya:2009uu}: our initial data is different, and the fact that we use different coordinate
systems makes direct comparisons somewhat involved.  Nonetheless, there are many similarities.  Our Gaussian profile falls off
so fast that it might as well be of compact support, and our choice of ingoing waves means that at large $r$ the profile is
essentially a function of an ingoing null coordinate $v$.  We can also choose our parameters $A$ and $\sigma$ (corresponding
respectively to the $\epsilon$ and $T$ of \cite{Bhattacharyya:2009uu}) sufficiently small to be within the perturbative
regime of \cite{Bhattacharyya:2009uu} or sufficiently large that that perturbative regime is no longer valid.
Furthermore, in both cases many of the main features of the collapse
process seem to depend simply on the approximate propagation of the scalar field along null geodesics until the shell becomes
sufficiently small that a trapped surface can form.
The main advantage of our method is that we can state rather precisely such things as
when a black hole forms and how large it is without having
to be in the regime where the perturbation expansion of \cite{Bhattacharyya:2009uu} is well approximated by its first
couple of terms.

To make contact with the Vaidya form of the metric we consider, in the metric \ref{eq:metric}, the following coordinate transformation:

\be
t=\mu(r,v), \qquad r=r.
\ee
The condition that the resulting metric has $v=\rm{const.}$ as ingoing null geodesics leads to
\be
\label{ingoing-null}
\frac{\partial \mu}{\partial r}=-\frac{a}{\alpha}.
\ee

We can adjust our metric explicitly to contain a null incoming coordinate:
\be
ds^2=-\alpha_v^2 dv^2+2\alpha_v a_v dv\, dr + r^2d\Omega^2,
\ee
where $\alpha_v=\left( \partial \mu/\partial v\right)\,\,\alpha$ and $a_v=a(r,v)$. One can now specialize the general form above to the particular
case of the Vaidya metric in global coordinates which is

\be
ds^2= -(1+\frac{r^2}{L^2}-\frac{M(v)}{r^2})dv^2 +2dr dv + r^2 d\Omega_3^2.
\ee
One important property of this metric which justifies is wide range of applications is that it solves Einstein's equation with {\it null} dust. Dust is
characterized only by its energy density which needs to be
\begin{equation}
T_{vv}=\frac{3}{2r^3} \frac{dM(v)}{dv}.
\end{equation}

In various recent works in the literature \cite{AbajoArrastia:2010yt,Aparicio:2011zy, Balasubramanian:2010ce,Balasubramanian:2011ur}
$M(v) \sim \tanh(v)$.  It is possible to look at figure (\ref{shell-norm-21})  and see how the collapse picture
differs from the thin shell model described by the Vaidya metric.  It is worth noticing that figure (\ref{shell-norm-21}) is rather representative of the hundred or so values of the width $\sigma$ that we have considered in this paper; only the particular details change but the qualitative picture is fairly universal.

%%%%%%%%%%%%%%%%%%%%%%%%%%%%%%%%%%%%%%%%%%%%%%%%%%%%%%%
\section{Comments on the dual field theory thermalization}\label{sec:ft}
%%%%%%%%%%%%%%%%%%%%%%%%%%%%%%%%%%%%%%%%%%%%%%%%%%%%%%%
%%%%%%%%%%%%%%%%%%%%%%%%%%%%%%%%%%%%%%%%%%%%%%%%%%%%%%%%%%%%%%%%%%%%%%%
{\subsection{Field theory setup}
%%%%%%%%%%%%%%%%%%%%%%%%%%%%%%%%%%%%%%%%%%%%%%%%%%%%%%%%%%%%%%%%%%%%%%%%
Our field theory setup is rather specific and we clarify it now. The overall field theoretic question one is interested in
answering is: How does a field theory react to a rapid injection of energy? This is precisely what the RHIC and LHC (ALICE) experiment are
all about for large values of the energy of the colliding particles -- How does QCD matter behave under such collision? Now
the collision is clearly anisotropic as one has two gold atoms colliding. Direct head-on collisions can potentially allow for semi isotropic treatments. A gravity approximation to the RHIC and ALICE experiment has
been developed in the context of numerical relativity by Chesler and Yaffe
in \cite{Chesler:2008hg,Chesler:2009cy,Chesler:2010bi}. Several authors have considered simpler versions of this problem by utilizing colliding shock waves but we do not discuss those works here. A natural time scale in such collision process is the
isotropization time which in the case of RHIC is $\tau_{iso}\le 1 fm/c$ \cite{Muller:2008zzm}.

It is worth pointing out that in the presence of an isotropization time scale we have to come to terms with
the existence of another scale which is the time scale where the hydrodynamic regime becomes applicable. Recall that
the thermalization time is, by definition, the time after which particles follow a Boltzmann distribution. The hydrodynamic
time, that is the time when the field theory is well described by the hydrodynamic approximation, can be applicable either
after the thermalization time or when the pressures in the rest frame become equal (isotropization). In the interesting
paper of the Chesler-Yaffe series \cite{Chesler:2010bi} evidence is found for the hydrodynamic regime to be applicable even
when the difference between the pressures in the longitudinal and transversal direction is still large, that is, the
hydrodynamic time scale is smaller than the isotropization time\footnote{We thank D. Mateos for this important
clarification.}. Note that other approaches, for example \cite{Heller:2011ju}, tend to identify the hydrodynamic
and isotropization times.

Our setup has one major difference. Namely, our injection of energy is spatially homogeneous, that is, it is the
same in all points of the field theory space at a given time. We thus study a situation that is similar but not
exactly equal to a quench. In this sense our setup is perhaps less directly applicable to RHIC-type experiments
of collision but rather speaks directly about more universal properties of strongly coupled field theories at
large $N$ which are dual to gravity theories. What we set out to study in this paper, using gravity methods, is
the time in field theory between the injection of energy at $t=0$  and the formation of the quark gluon plasma. We call
this time the thermalization time. Of course, other interesting time scales might be present in a given experimental
setup. One example is the above mentioned isotropization time, another natural time is the time at which the hydrodynamical
approximation becomes a valid description of the quark gluon plasma. The thermalization time, as defined in our context, is
expected to be of the same order but slightly smaller than these other two.

To conclude this part by emphasizing that there are various time scales in the problem some of which we list:

\begin{itemize}
\item Hydrodynamic time scale: scale after which the liquid is well described by hydrodynamics, small velocities, long wavelengths.
\item Isotropization time: The time scale after which the pressure differences in various directions is small.
\item Thermalization time: Time after which the particle distribution follows the Boltzmann distribution.

\end{itemize}

%

%%%%%%%%%%%%%%%%%%%%%%%%%%%%%%%%%%%%%%%%%%%%%%%%%%%%%%%%%%%%
\subsection{Field theory regimes}
%%%%%%%%%%%%%%%%%%%%%%%%%%%%%%%%%%%%%%%%%%%%%%%%%%%%%%%%%%%%%
As we know, in field theory not every collision or injection of energy will lead to formation of a thermal state. Experiments that involve low energies, and low number of particles will be better described as scattering states. As the energy and number of particles involved increases we expect the final state to be better approximated by a thermal state. As demonstrated in the RHIC experiment, at strong coupling the energy needed to generate quark-gluon plasma is lower than the naively  expected, however still finite.

Recent results due to Bizon and collaborators \cite{Bizon:2011gg,Jalmuzna:2011qw} and also corroborated in \cite{Dias:2011ss} indicate that any perturbation in AdS ultimately leads to gravitational collapse. We would like to explain how these results are reconciled with the standard field theory expectations.

From the gravity point of view Bizon's result translates into a formation of a thermal state for a collision of arbitrary small energies or for very small amount of energy. However, when translated into a field theory language we need to incorporate the large $N_c$ scaling. The energy scale for the collapsing shell involve Newton's constant $\kappa^2$, when properly translated into field theory language we counter a $N_c^2$ scaling of all energies, equivalently an order of $N_c^2$ particles. In terms of the AdS/CFT duality the gravity approximation is valid only for infinite $N_c$, thus in term of field theory we always have enough particles to form a proper thermal state. If we were to discuss field theory collisions that are too small to generate quark-qluon plasma, we should introduce stringy objects and quantum excitations in the AdS side.

One can summarize the above discussion by identifying two regimes:\\

\begin{itemize}

\item  Regime without backreaction: The field theory description of a collision with very low energy or a low energy quench
should find its gravity description in the geodesic motion of objects in $AdS$.\\

\item Regime with backreaction: The formation of a black hole with nonzero area in the gravity side is equivalent to the formation
of a quark gluon plasma in the field theory. The fact that this always happens when considering a scalar field in Einstein gravity
with a negative cosmological constant is  a consequence of the large $N_c$ limit. We will show more rigorously in the next section that this scaling does indeed take place.

\end{itemize}

%%%%%%%%%%%%%%%%%%%%%%%%%%%%%%%%%%%%%%%%%%%%%%%%%%%%%%%
\subsection{Hawking-Page transition and gravitational collapse: Canonical and microcanonical ensembles}\label{sec:HP}
%%%%%%%%%%%%%%%%%%%%%%%%%%%%%%%%%%%%%%%%%%%%%%%%%%%%%%%
There are many arguments leading to the fact that the gravity calculations presented here pertain to the microcanonical ensemble. This questions has been revisited several times in the context of general relativity. First, conservation of energy tells us that we are in a system with a fixed amount of total energy. Further, since black holes in flat space have negative specific heat they can cool themselves by transferring energy to the surrounding and consequently be able to transfer more energy. We expect that small black holes in $AdS$ will behave similarly.

From the field theory point of view sometimes the canonical ensemble is relevant in which case it is worth revisiting the role of the Hawking-Page phase transition in the context of the AdS/CFT correspondence.  The interpretation of the Hawking-Page transition in the context of the gauge gravity correspondence leads to the result that below a certain temperature, that is, below a certain radius of the black hole, the background that is relevant in the
thermodynamic competition is no longer the Schwarzschild black hole but rather thermal $AdS$ \cite{Witten:1998zw}. This result
means that for low temperatures the field theory goes into a confined phase rather than remaining in the quark gluon
plasma. These suggests that the very low temperature black holes are not relevant from the field theory point of
view as they correspond to an unstable phase.

The Hawking-Page transition occurs, using Witten's notation \cite{Witten:1998zw}, at  $r_+=L$. Namely
\be
I=\frac{{\rm vol}(S^{n-1})}{4G_N} \frac{ L^2 r_{+}^{n-1}-r_+^{n+1}}{nr_+^2+(n-2)L^2}
\ee
For $r_+<L$ the thermodynamical competition is won by thermal $AdS_{n+1}$; for $r_+>L$ the dominant themodynamical phase is
the black hole. In the work \cite{Garfinkle:2011hm} and the discussions of \cite{Bizon:2011gg,Jalmuzna:2011qw}, black holes with $r_+<L=1$ should
not be considered as part of the description of a dual field theory since in that range the field theory stable phase is best described by thermal $AdS$. A counting of the degrees of freedoms allows to identify this transition with the confinement/deconfinement transition given that the entropy of the black hole is of the order $N_c^2$ while the entropy of thermal $AdS$ is of the order $N_c^0$.

%%%%%%%%%%%%%%%%%%%%%%%%%%%%%%%%%%%%%%%%%%%%%%%%%%%%%%%%%%%%
\subsection{Boundary stress energy tensor}
%%%%%%%%%%%%%%%%%%%%%%%%%%%%%%%%%%%%%%%%%%%%%%%%%%%%%%%%%%%%%
Translating gravity quantities into field theory terms is, in general, a task that has to be dealt with on a case by case basis. In this section we will simply use the dictionary for the canonical case of the duality in the context of classical strings in $AdS_5\times S^5$ dual to  ${\cal N}=4$ SYM:
\be
\frac{1}{\kappa^2}= \frac{L^5}{64\pi^4g_s^2 \alpha'{}^4}, \qquad L^4= 4\pi g_s \alpha'{}^2.
\ee
For many quantities, for example Kubo-type formulas, one can verify that the gravity quantities combine in a way
that there is no dependence in $\alpha'$ leading to
\be
\frac{L^3}{\kappa^2}=\frac{N_c^2}{4\pi^2}.
\ee
Our goal in this section is to extract the field theory stress energy tensor. The prescription for extracting the boundary stress energy tensor using bulk gravity data is well established in the context of the AdS/CFT correspondence  \cite{Balasubramanian:1999re}\cite{deHaro:2000xn} and reads:

\bea
\langle T_{\mu\nu} \rangle &=& \frac{N_c^2}{4\pi^2} \lim\limits_{r\to \infty}r^4 \delta g_{\mu\nu}
\eea
where $\mu, \nu$ run over the field theory directions and $\delta g_{\mu\nu}$ is defined as the difference of the resulting metric from the boundary metric corresponding to $AdS$. We are interested in the boundary or field theory energy which we relate to $T_{00}$ in the sense above.  We are thus, interested in studying the asymptotic behavior of the function $\alpha(t,r)$ from which we read $\delta g_{00}$ above. The most efficient approach is to approximate, in the asymptotic region, the function $\alpha$ by its Schwarzschild-AdS expression. Using the equations of motion we could, equally effectively study how well approximated is the function $a$ by its Schwarzschild-AdS expression which is what we actually do in the simulations.

Figure \ref{Fig:asymptotics} shows how well the function $a(r,t)$ is approximated by the Schwarzschild-AdS expression at times close to the time of formation of an apparent horizon. To aid understanding the fit better we have plotted the difference between the function $a(r,t)$ and its Schwarzschild-AdS expression. It is remarkable that this particular values of the simulation parameters the difference is better than $10^{-8}$ as we approach the asymptotic regions.
\begin{figure}[htp]
\begin{center}
\includegraphics[width=4.0in]{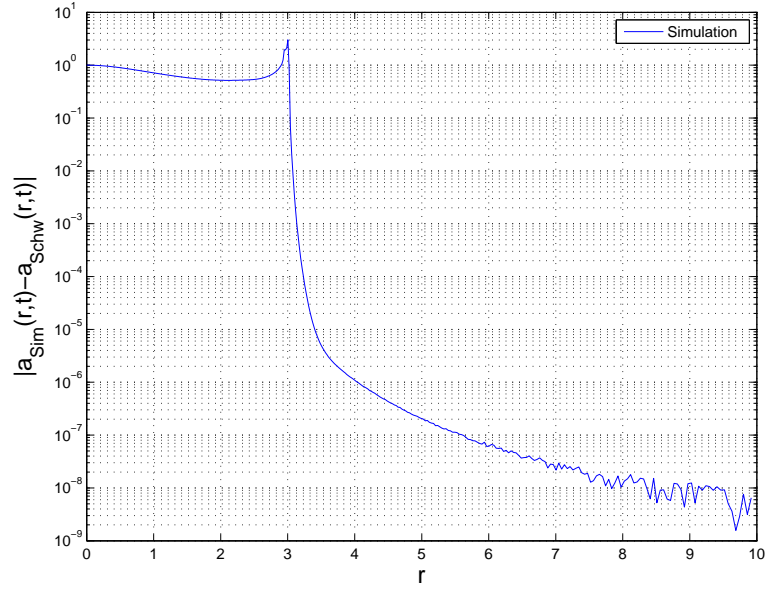}
\caption{\label{Fig:asymptotics} The difference close to $t_{\rm AH}$ of the metric function  $a(r,t\approx t_{AH})$ directly from the simulation and a fit with the static Schwarzschild-AdS metric. }
\end{center}
\end{figure}

Finally we consider the same plot for a hundred different values of the amplitude of the scalar field. The fitting is better than half a percent for the majority of the values tested. Except for the two anomalous points, larger amplitudes yield a better fitting to the Schwarzschild-AdS form of the metric. Notice that at small horizon radius we do expect a deviation between $r_{BH}$ and $r_{AH}$ as at the time of formation of the trapped surface, some portion of the scalar energy is still un-trapped.

\begin{figure}[htp]
\begin{center}
\includegraphics[width=4.0in]{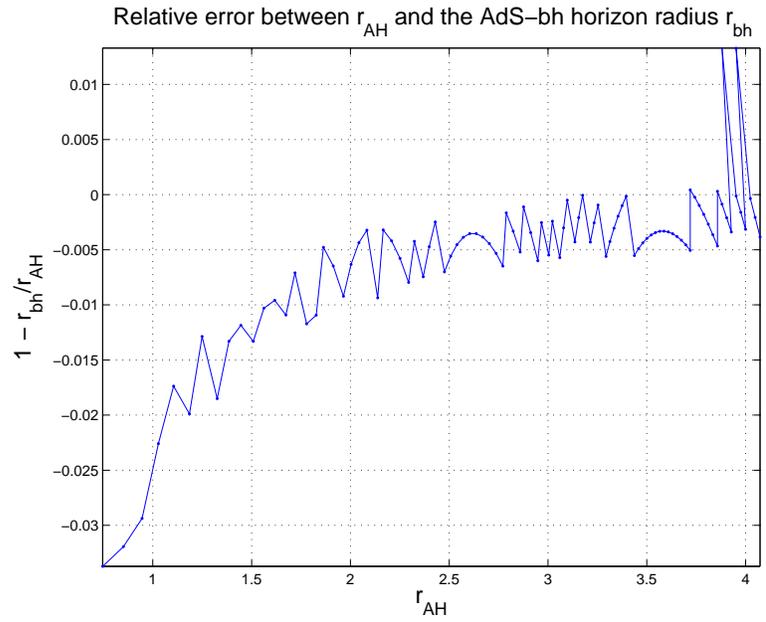}
\caption{\label{relative-error} Relative error in determining the horizon radius from the simulation value of the apparent horizon and from the fit of the metric functions according to the static metric of Schwarzschild-AdS }
\end{center}
\end{figure}

%%%%%%%%%%%%%%%%%%%%%%%%%%%%%%%%%%%%%%%%%%%%%%%%%%%%%%%%%%%%
\subsection{RHIC and LHC isotropization times}
%%%%%%%%%%%%%%%%%%%%%%%%%%%%%%%%%%%%%%%%%%%%%%%%%%%%%%%%%%%%%
In this section, despite all the disclaimers made previously, we take our results in the direction of application to actual experiments. Let us remark again that we understand that thermalization time  and $t_{AH}$ could be widely different. Our goal is not to make a concrete numerical prediction, rather, we aim at simply illustrating where our results  might fit. One of the results of the gravity simulation is a description of the dependence of the time of formation of an apparent horizon as a function of the black hole mass. When translated into field theory terms we could argue that this graph approximates the dependence of the thermalization time (formation of the quark gluon plasma roughly) as a function of the energy. Below, in figure \ref{TvM},  we plot such graph.

\begin{figure}[htp]
\begin{center}
\includegraphics[width=4.0in]{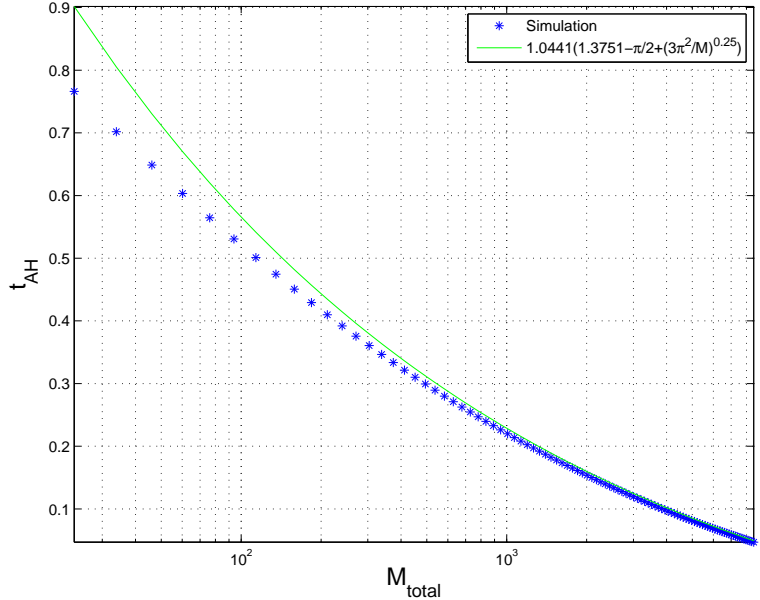}
\caption{\label{TvM} The dependence of $t_{\rm AH}$ on the mass of the formed black hole }
\end{center}
\end{figure}

The best fit to the curve in figure (\ref{TvM}) could be understood as a curve describing $t_{Thermalization}(E )$. We shall loosely interpret $E$ as the energy injected into the field theory. We consider explicitly a fit for large masses which clearly follows from the geodesic approximation presented in fig. (\ref{TvM}). Note that this expression is a deviation of the naive one that can be obtained in the null geodesic approximation for $AdS$. Namely, from equation (\ref{Eq:t_AH}) it follows, in the large mass limit there, that $t\approx E^{-0.25}$ where we relate $E$ to the radius of the black hole using the Schwarzschild solution. The precise numbers appearing in fig.  (\ref{TvM}) have the same interpretation discussed in section \ref{sec:bhformation}. Without providing further evidence we mention that we have also considered similar graphs for different widths of the scalar field and the results come out in the same range of the exponential. We do not claim that this is a universal behavior but within our modest context this exponent has come out about the same for different values of the widths. It will be interesting to explore this potential universality in the $t_{AH}$ versus $M_{BH}$ graph further.

Hoping that the analysis we performed is universal enough we dare to compare our results with experimental results.  There
is really no reason this might work as we are comparing times in very different experiments. As explained in the introduction
the thermalization time we compute here and the isotropization time at RHIC and LHC correspond to different physical
situations. However, there is the possibility that the time we compute here is of the same order as the isotropization
time as it describes a fairly universal property of the dual field theory, namely, its response to a a sudden influx of
energy. With that disclaimer, we could conclude that $\frac{\tau_{LHC}}{\tau_{RHIC}} =\frac{f(E_{RHIC})}{f(E_{LHC})}$. We use \cite{Back:2004je} to estimate that the central  Au-Au collisions occur at approximately  $\sqrt{s_{NN}}=0.2 TeV$ while the data of
the ALICE collaboration at CERN \cite{Aamodt:2010pa}, \cite{Aamodt:2010pb} suggest that the central Pb-Pb collisions at $\sqrt{s_{NN}}=2.76 TeV$. Plugging in the numbers we get $\tau_{LHC}\approx .5 \tau_{RHIC}$.

%%%%%%%%%%%%%%%%%%%%%%%%%%%%%%%%%%%%%%%%%%%%%%%%%%%%%%%
\section{Conclusions}\label{sec:Conclusions}
%%%%%%%%%%%%%%%%%%%%%%%%%%%%%%%%%%%%%%%%%%%%%%%%%%%%%%%
In this paper we have considered gravitational collapse of a scalar field coupled to Einstein gravity with a negative cosmological constant. We have focused on the dependence of the time of formation of an apparent horizon as a function of the parameters characterizing an initial Gaussian profile for the scalar field. We varied the width and amplitude of the initial Gaussian profile and studied how it affects the time of formation of an apparent horizon. We obtained results for masses ranging over two orders
of magnitude. In a more detailed study we verify that the dependence on the width is milder than the dependence on the amplitude. Our simulations indicate a saturation of sorts, in the time of formation of an apparent horizon, as we increase the width of the Gaussian profile. In conjunction with \cite{Garfinkle:2011hm} we continue to build on the foundation that will allow us to provide, in full detail, various other properties of the thermalization process by studying its
gravitational dual.  We have proposed that these studies might provide rough estimates for quantities that depend on the thermalization time. Even though the concept of thermalization itself  depends on the precise set up, gravity studies should provide a rough estimate.

We conclude enumerating some of the directions that we believe would be interesting to pursue.

\begin{itemize}

\item We have verified that for some parameters in the simulations the scalar field does collapse after a few bounces as explained in  \cite{Bizon:2011gg} and \cite{Jalmuzna:2011qw}. The relevance of this result for the microcanonical ensemble in the field theory dual is hard to overstate and should be clarified.

\item What is the role of the potential in the scalar field? It is our expectation that considering a potential term of the form $m^2\phi^2$ could significantly impact our results as an underlying intuition throughout the paper has been the behavior of null geodesics.

\item Spherically symmetric approximation is poor for describing collisions. Key properties, such as elliptic flow, are missing. Although the techniques involved in going beyond the spherically symmetric framework are substantial it is worth pursuing.

\item Other interesting time-dependent phenomena have recently been tackled using gauge/gravity correspondence. In particular
aging phenomena have been discussed in \cite{Minic:2008xa,Nakayama:2010xq,Jottar:2010vp}. It would be nice to have a more precise
gravity description along the lines obtained in the context of this time-dependent far-from-equilibrium phenomena, aging. Note
that \cite{Jottar:2010vp} obtained the field theory correlators using a gravity background and the corresponding gauge/gravity
prescription for correlators. One really intriguing question would be to understand precisely what initial quenches lead to
black hole formation and which lead to an aging situation. A numerical realization along  the lines presented in this paper of
aging phenomena would be a great insight into the structure of aging phenomena.

\item Finally and perhaps most interesting conceptually: Can we use our solutions to set up a laboratory to study the black hole information paradox?  There is no doubt that the information paradox is resolved in the context of the gauge/gravity
correspondence as we have a map between a gravity theory with black hole formation and evaporation and a unitary field
theory. The answer to the information paradox has rather become a question of {\it how} it gets resolved in our models of gravitational collapse. A detailed study of black hole formation with the various stringy probes and correlation functions is a tantalizing route to understanding this question better and to an eventual explicit resolution of some of the most interesting black hole puzzles.

\end{itemize}

%%%%%%%%%%%%%%%%%%%%%%%%%%%%%%%%%%%%%%%%%%%%%%%%%%%%%%%
\section*{Acknowledgments}
%%%%%%%%%%%%%%%%%%%%%%%%%%%%%%%%%%%%%%%%%%%%%%%%%%%%%%%
L. PZ is grateful to H. de Oliveira and C. Terrero-Escalante for collaboration on similar matters. We thank P. Biz\'on, J. Erdmenger, G. Horowitz, S. Khlebnikov,  B. Kol, M. Kruczenski, L. Lehner, E, L\'opez, D. Mateos, S. Minwalla and D. Trancanelli for comments. L. PZ is thankful to E. Rasia for
many patient explanations on Fortran. This work is  partially supported by Department of Energy under grant
DE-FG02-95ER40899 to the University of Michigan and by NSF grant PHY-0855532 to Oakland University.
\newpage
\begin{appendices}

%%%%%%%%%%%%%%%%%%%%%%%%%%%%%%%%%%%%%%%%%%%%%%%%%%%%%%%
\section{Equations of Motion}\label{app:EoM}
%%%%%%%%%%%%%%%%%%%%%%%%%%%%%%%%%%%%%%%%%%%%%%%%%%%%%%%

The Einstein tensor $G_{\mu\nu}=R_{\mu\nu}-\frac12 g_{\mu\nu} R$ is
\bea
G_{tt}&=&\frac{3\alpha^2 }{r^2\, a^3}\bigg[-a(1-a^2)+r\partial_r a \bigg], \nonumber \\
G_{tr}&=& \frac{3\partial_t a}{r\, a}, \nonumber \\
G_{rr}&=& \frac{3}{r^2\, \alpha}\bigg[(1-a^2)\alpha + r \partial_r \alpha\bigg],\nonumber \\
G_{\chi\chi}&=&\frac{1}{a^3\, \alpha^3}\bigg[-a^3\alpha^3 - r \alpha^2 \partial_r a (2\alpha + \partial_r a)+a\alpha^2 (\alpha + r (2 \partial_r \alpha +r \partial_{r}^2 \alpha ) + r^2 a^2 (\partial_t a \partial_t \alpha -\alpha \partial^2_t a)\bigg], \nonumber \\
G_{\theta\theta}&=&G_{\chi\chi}\sin^2 \chi, \nonumber \\
G_{\varphi\varphi}&=&G_{\chi\chi}\sin^2 \chi\sin^2 \theta.
\eea
The stress energy tensor is written in a form that contains the contribution from the cosmological constant
\bea
T_{\mu\nu}&=&\partial_\mu \phi \partial_\nu \phi + g_{\mu\nu} \left(\frac{6}{L^2}-\frac12 (\partial \phi )^2 -U(\varphi)\right), \nonumber \\
T_{tt}&=& -\frac{6\alpha^2}{L^2}+(\partial_t\phi)^2+\frac{\alpha^2}{2}((\partial\phi)^2 +2U(\phi)), \nonumber \\
T_{tr}&=&\partial_t\phi \partial_r \phi, \nonumber \\
T_{rr}&=& \frac{6 \,a^2}{L^2}+(\partial_r\phi)^2 -\frac{a^2}{2} ((\partial\phi)^2 +2U(\phi)), \nonumber \\
T_{\chi\chi}&=&\frac{6\,r^2}{L^2}-\frac{r^2}{2} \left((\partial \phi )^2 +2U(\phi)\right), \nonumber \\
\eea

%%%%%%%%%%%%%%%%%%%%%%%%%%%%%%%%%%%%%%%%%%%%%%%%%%%%%%%%%%%%%%%%%%%%%%%%%%%%%%%%%%%%%%%%%%%%%%
\section{Geodesics in global $AdS_5$}\label{app:geodesics}
%%%%%%%%%%%%%%%%%%%%%%%%%%%%%%%%%%%%%%%%%%%%%%%%%%%%%%%%%%%%%%%%%%%%%%%%%%%%%%%%%%%%%%%%%%%%%%%%
In this appendix we recall a few known facts about $AdS$ in global coordinates.
We discuss a massless scalar field that should move along a null geodesic. Recall that a null geodesic reaches spatial
infinity in finite asymptotic time.  Namely, considering the metric
\be
\label{eq:metricAdS}
ds^2 = -\left(1+\frac{r^2}{L^2}\right)dt^2 + \frac{dr^2}{1+\frac{r^2}{L^2}}+r^2 d\Omega_3^2.
\ee
For a radial {\it null} geodic we have
\be
\frac{dr}{d\lambda}=\pm E, \qquad \frac{dt}{d\lambda}=\frac{E}{1+\frac{r^2}{L^2}}.
\ee
In particular such a null geodesic reaches $r\to \infty$ starting at $r=0$ in
\be
t=\int\limits_0^\infty \frac{dr}{1+\frac{r^2}{L^2}}=\frac{\pi}{2}L.
\ee
In our analysis of gravitational collapse we will naturally observe that for solutions where the amplitude of the scalar
field is not sufficient to form a black hole the wave will roughly oscillate many times  which is a combination
of reaching asymptotic infinity in finite coordinate time and the fact that we impose Dirichlet boundary conditions at
infinity since AdS is not globally hyperbolic.

The situation for massive geodesics is similar. Starting from the metric (\ref{eq:metricAdS}), the relevant equations of motion are
\be
\frac{dt}{d\tau}=\frac{E}{1+\frac{r^2}{L^2}}, \qquad \dot{r}^2 +\frac{r^2}{L^2}=E^2-1.
\ee
The solution of this massive geodesic is
\be
r(\lambda)=L \sqrt{E^2-1} \sin \left(\frac{1}{L}\lambda\right).
\ee
Interestingly, the time that it takes this particle to reach its maximum radius $r_{max}=L\sqrt{E^2-1}$ is precisely as
in the case of the null particle
\be
t=\int\limits_0^{r_{max}}dr \frac{E}{(1+\frac{r^2}{L^2})\sqrt{E^2-1-\frac{r^2}{L^2}}}=\frac{\pi}{2}\, L.
\ee

%%%%%%%%%%%%%%%%%%%%%%%%%%%%%%%%%%%%%%%%%%%%%%%%%%%%%%%%%%%%%%%%%%%%%%%%%%%%%%%%%%%%%%%%%%%%%55
\section{Times: simulation, coordinate, physical and field theory times}\label{app:times}

In this appendix we clarify the different measures of time that we used throughout the paper.
The coordinate time is simply $t$. We make sure that at each step the simulation time step $dt$ is at least smaller than the
corresponding physical time step associate with an increment in the radial direction by $dr$, that is, $dt \le \left( a/\alpha\right) dr$. In the  bulk of the paper we use this coordinate or simulation time and call it $t_{Sim}$.

The metric given in equation (\ref{eq:metric}) has still a freedom that we would like to fix. Namely, the form of the metric is unchanged if we apply
a transformation of the form $t\to f(t), \alpha \to \alpha f'(t)$. We would like to use a gauge in which the time at infinity is normalized in the
standard form. However, that was not the choice we took in the simulations, rather we used $\alpha(r=0,t)=1$. Because of  this choice we use
in some of the graphs what we call field theory time
\be
dt_{FT}=a(r=r_{max},t_{Sim}) \, \alpha(r=r_{max},t_{Sim})  dt_{Sim}.
\ee
In the beginning of the simulation the difference between $t_{Sim}$ and $t_{FT}$ is essentially a constant that accounts for the right field theory
normalization. However, near the time of formation of the apparent horizon the situation is quite different as can be seen from the plots in figure (\ref{tSimVtFT}). Note that that figure (\ref{tSimVtFT}) provides a check of the stability of our definition as a function of the grid resolution. Namely, despite the violent dependence around the formation of the apparent horizon different resolutions behave remarkably close to each other are are essentially indistinguishable in the plots. Since the relation between the field theory time and the simulation time is via an integration, figure (\ref{tSimVtFT}) shows that the accumulated error is negligible among the various grid resolutions that we used. However, near the time of formation of the apparent horizon $t_{FT}$ breaks down and should not be used to define field theory thermalization time.

\begin{figure}
\begin{center}
\includegraphics[width=4.0in]{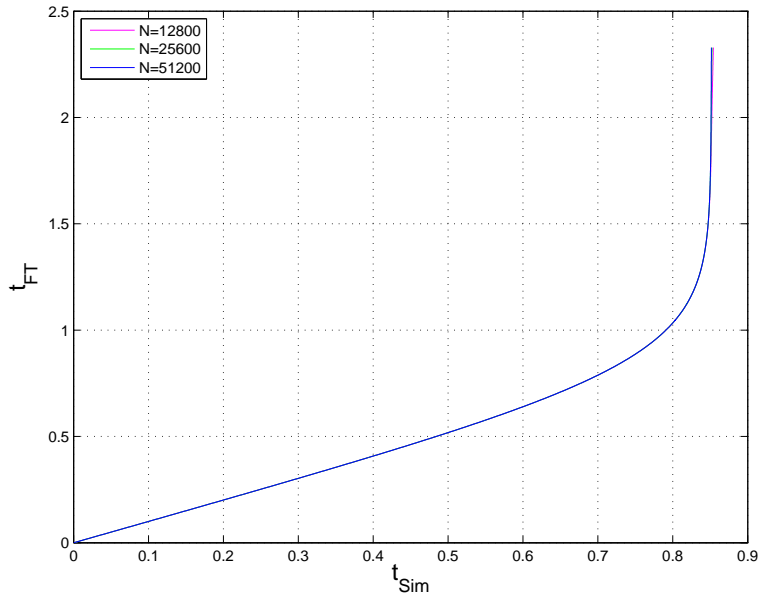}
\caption{\label{tSimVtFT} Plot of the field theory time $t_{FT}$ versus the coordinate or simulation time $t_{Sim}$. }
\end{center}
\end{figure}

To develop a better intuition of the implications of using field theory time, we consider the constraint plotted as a function of the field theory time. Notice that the constraints have a  milder behavior near the time of horizon formation as
can be seen in the figure (\ref{L2NormstFT}). This figure should be compared directly with figure (\ref{L2Norms}) of section \ref{sec:accurary}. The key feature is the smoothing around the time of formation of the apparent horizon which indicates that the simulations are not really losing accuracy as one could infer from figure (\ref{L2Norms}) of section \ref{sec:accurary}.

\begin{figure}
\begin{center}
\includegraphics[width=4.0in]{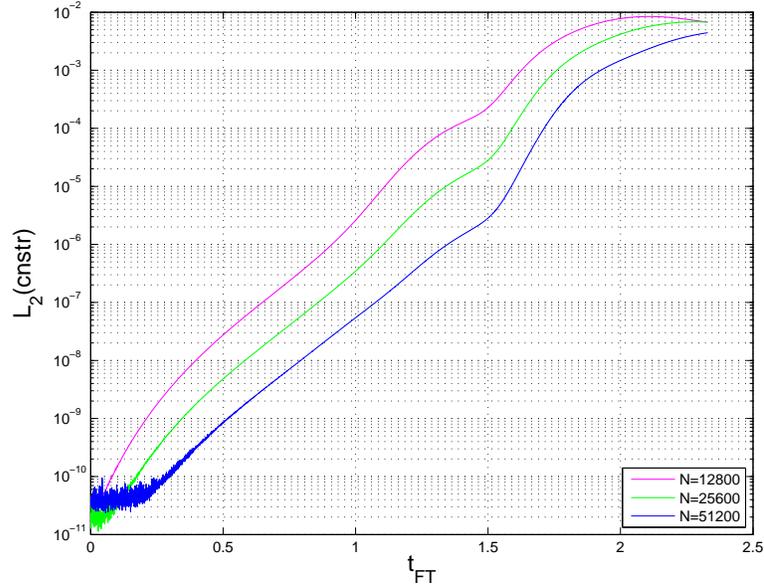}
\caption{\label{L2NormstFT} The $L_2$ norm of the constraint for various values of the resolution as a function of the field theory time. }
\end{center}
\end{figure}

\end{appendices}

Motivated by the remarks of section (\ref{sec:Comparing})  we attempted a computation of the ingoing null geodesic time denoted by $v$. This coordinate is essentially what would be required to go to the Eddington-Finkelstein coordinates. This null time $v$ is a key ingredient in various AdS/CFT setups, certainly in computations of spectral functions it is this time that naturally fixes the boundary conditions near the horizon. It also figures prominently in the hydrodynamic context (see \cite{Son:2007vk,Hubeny:2010ry,Hubeny:2011hd}). The setup used in the bulk of the paper does not lead to a well defined $v$.

\newpage
\newpage
% If you don't have the corresponding .bst and .bib files, comment the two lines below and copy paste the content of the .bbl file
%there (whoever compiled the %bibliography should send you the .bbl file)
\bibliographystyle{JHEP}
\bibliography{Collapsebib}

\providecommand{\href}[2]{#2}\begingroup\raggedright\begin{thebibliography}{10}

\bibitem{Maldacena:1997re}
J.~M. Maldacena, {\it {The large N limit of superconformal field theories and
  supergravity}},  {\em Adv. Theor. Math. Phys.} {\bf 2} (1998) 231--252,
  [\href{http://xxx.lanl.gov/abs/hep-th/9711200}{{\tt hep-th/9711200}}].

\bibitem{Witten:1998qj}
E.~Witten, {\it {Anti-de Sitter space and holography}},  {\em Adv. Theor. Math.
  Phys.} {\bf 2} (1998) 253--291,
  [\href{http://xxx.lanl.gov/abs/hep-th/9802150}{{\tt hep-th/9802150}}].

\bibitem{Gubser:1998bc}
S.~S. Gubser, I.~R. Klebanov, and A.~M. Polyakov, {\it {Gauge theory
  correlators from non-critical string theory}},  {\em Phys. Lett.} {\bf B428}
  (1998) 105--114, [\href{http://xxx.lanl.gov/abs/hep-th/9802109}{{\tt
  hep-th/9802109}}].

\bibitem{Aharony:1999ti}
O.~Aharony, S.~S. Gubser, J.~M. Maldacena, H.~Ooguri, and Y.~Oz, {\it {Large N
  field theories, string theory and gravity}},  {\em Phys. Rept.} {\bf 323}
  (2000) 183--386, [\href{http://xxx.lanl.gov/abs/hep-th/9905111}{{\tt
  hep-th/9905111}}].

\bibitem{Witten:1998zw}
E.~Witten, {\it {Anti-de Sitter space, thermal phase transition, and
  confinement in gauge theories}},  {\em Adv. Theor. Math. Phys.} {\bf 2}
  (1998) 505--532, [\href{http://xxx.lanl.gov/abs/hep-th/9803131}{{\tt
  hep-th/9803131}}].

\bibitem{Son:2007vk}
D.~T. Son and A.~O. Starinets, {\it {Viscosity, Black Holes, and Quantum Field
  Theory}},  {\em Ann. Rev. Nucl. Part. Sci.} {\bf 57} (2007) 95--118,
  [\href{http://xxx.lanl.gov/abs/0704.0240}{{\tt arXiv:0704.0240}}].

\bibitem{Hubeny:2010ry}
V.~E. Hubeny and M.~Rangamani, {\it {A Holographic view on physics out of
  equilibrium}},  {\em Adv.High Energy Phys.} {\bf 2010} (2010) 297916,
  [\href{http://xxx.lanl.gov/abs/1006.3675}{{\tt arXiv:1006.3675}}].

\bibitem{Hubeny:2011hd}
V.~E. Hubeny, S.~Minwalla, and M.~Rangamani, {\it {The fluid/gravity
  correspondence}},  \href{http://xxx.lanl.gov/abs/1107.5780}{{\tt
  arXiv:1107.5780}}. * Temporary entry *.

\bibitem{Shuryak:2003xe}
E.~Shuryak, {\it {Why does the quark gluon plasma at RHIC behave as a nearly
  ideal fluid?}},  {\em Prog. Part. Nucl. Phys.} {\bf 53} (2004) 273--303,
  [\href{http://xxx.lanl.gov/abs/hep-ph/0312227}{{\tt hep-ph/0312227}}].

\bibitem{Shuryak:2004cy}
E.~V. Shuryak, {\it {What RHIC experiments and theory tell us about properties
  of quark-gluon plasma?}},  {\em Nucl. Phys.} {\bf A750} (2005) 64--83,
  [\href{http://xxx.lanl.gov/abs/hep-ph/0405066}{{\tt hep-ph/0405066}}].

\bibitem{Heinz:2004pj}
U.~W. Heinz, {\it {Thermalization at RHIC}},  {\em AIP Conf. Proc.} {\bf 739}
  (2005) 163--180, [\href{http://xxx.lanl.gov/abs/nucl-th/0407067}{{\tt
  nucl-th/0407067}}].

\bibitem{Muller:2008zzm}
B.~Muller, {\it {Theoretical challenges posed by the data from RHIC}},  {\em
  Prog. Theor. Phys. Suppl.} {\bf 174} (2008) 103--121.

\bibitem{Garfinkle:2011hm}
D.~Garfinkle and L.~A. Pando~Zayas, {\it {Rapid Thermalization in Field Theory
  from Gravitational Collapse}},  {\em Phys.Rev.} {\bf D84} (2011) 066006,
  [\href{http://xxx.lanl.gov/abs/1106.2339}{{\tt arXiv:1106.2339}}].

\bibitem{Hartnoll:2008vx}
S.~A. Hartnoll, C.~P. Herzog, and G.~T. Horowitz, {\it {Building a Holographic
  Superconductor}},  {\em Phys. Rev. Lett.} {\bf 101} (2008) 031601,
  [\href{http://xxx.lanl.gov/abs/0803.3295}{{\tt arXiv:0803.3295}}].

\bibitem{Gubser:2008zu}
S.~S. Gubser, {\it {Colorful horizons with charge in anti-de Sitter space}},
  {\em Phys. Rev. Lett.} {\bf 101} (2008) 191601,
  [\href{http://xxx.lanl.gov/abs/0803.3483}{{\tt arXiv:0803.3483}}].

\bibitem{Hartnoll:2009sz}
S.~A. Hartnoll, {\it {Lectures on holographic methods for condensed matter
  physics}},  {\em Class.Quant.Grav.} {\bf 26} (2009) 224002,
  [\href{http://xxx.lanl.gov/abs/0903.3246}{{\tt arXiv:0903.3246}}].

\bibitem{Horowitz:2010gk}
G.~T. Horowitz, {\it {Introduction to Holographic Superconductors}},
  \href{http://xxx.lanl.gov/abs/1002.1722}{{\tt arXiv:1002.1722}}.

\bibitem{Cubrovic:2009ye}
M.~Cubrovic, J.~Zaanen, and K.~Schalm, {\it {String Theory, Quantum Phase
  Transitions and the Emergent Fermi-Liquid}},  {\em Science} {\bf 325} (2009)
  439--444, [\href{http://xxx.lanl.gov/abs/0904.1993}{{\tt arXiv:0904.1993}}].

\bibitem{Faulkner:2010zz}
T.~Faulkner, N.~Iqbal, H.~Liu, J.~McGreevy, and D.~Vegh, {\it {Strange metal
  transport realized by gauge/gravity duality}},  {\em Science} {\bf 329}
  (2010) 1043--1047.

\bibitem{McGreevy:2009xe}
J.~McGreevy, {\it {Holographic duality with a view toward many-body physics}},
  {\em Adv.High Energy Phys.} {\bf 2010} (2010) 723105,
  [\href{http://xxx.lanl.gov/abs/0909.0518}{{\tt arXiv:0909.0518}}].

\bibitem{RigolNature}
M.~Rigol, V.~Dunjko, and M.~Olshanii, {\it Thermalization and its mechanism for
  generic isolated quantum systems},  {\em Nature} {\bf 452} (2008) 854--858.

\bibitem{PhysRevLett.103.100403}
M.~Rigol, {\it Breakdown of thermalization in finite one-dimensional systems},
  {\em Phys. Rev. Lett.} {\bf 103} (Sep, 2009) 100403.

\bibitem{Calabrese:2006rx}
P.~Calabrese and J.~L. Cardy, {\it {Time-dependence of correlation functions
  following a quantum quench}},  {\em Phys. Rev. Lett.} {\bf 96} (2006) 136801,
  [\href{http://xxx.lanl.gov/abs/cond-mat/0601225}{{\tt cond-mat/0601225}}].

\bibitem{Cardy:2011zz}
J.~Cardy, {\it {Measuring entanglement using quantum quenches}},  {\em Phys.
  Rev. Lett.} {\bf 106} (2011) 150404,
  [\href{http://xxx.lanl.gov/abs/1012.5116}{{\tt arXiv:1012.5116}}].

\bibitem{AbajoArrastia:2010yt}
J.~Abajo-Arrastia, J.~Aparicio, and E.~Lopez, {\it {Holographic Evolution of
  Entanglement Entropy}},  {\em JHEP} {\bf 11} (2010) 149,
  [\href{http://xxx.lanl.gov/abs/1006.4090}{{\tt arXiv:1006.4090}}].

\bibitem{Aparicio:2011zy}
J.~Aparicio and E.~Lopez, {\it {Evolution of Two-Point Functions from
  Holography}},  \href{http://xxx.lanl.gov/abs/1109.3571}{{\tt
  arXiv:1109.3571}}. * Temporary entry *.

\bibitem{PhysRevLett.88.160401}
Z.~Hadzibabic, C.~A. Stan, K.~Dieckmann, S.~Gupta, M.~W. Zwierlein,
  A.~G\"orlitz, and W.~Ketterle, {\it Two-species mixture of quantum degenerate
  bose and fermi gases},  {\em Phys. Rev. Lett.} {\bf 88} (Apr, 2002) 160401.

\bibitem{Bhattacharyya:2009uu}
S.~Bhattacharyya and S.~Minwalla, {\it {Weak Field Black Hole Formation in
  Asymptotically AdS Spacetimes}},  {\em JHEP} {\bf 0909} (2009) 034,
  [\href{http://xxx.lanl.gov/abs/0904.0464}{{\tt arXiv:0904.0464}}].

\bibitem{Balasubramanian:2010ce}
V.~Balasubramanian, A.~Bernamonti, J.~de~Boer, N.~Copland, B.~Craps, {\em
  et.~al.}, {\it {Thermalization of Strongly Coupled Field Theories}},
  \href{http://xxx.lanl.gov/abs/1012.4753}{{\tt arXiv:1012.4753}}. * Temporary
  entry *.

\bibitem{Erdmenger:2011jb}
J.~Erdmenger, S.~Lin, and T.~H. Ngo, {\it {A moving mirror in AdS space as a
  toy model for holographic thermalization}},  {\em JHEP} {\bf 04} (2011) 035,
  [\href{http://xxx.lanl.gov/abs/1101.5505}{{\tt arXiv:1101.5505}}].

\bibitem{Balasubramanian:2011ur}
V.~Balasubramanian, A.~Bernamonti, J.~de~Boer, N.~B. Copland, B.~Craps, {\em
  et.~al.}, {\it {Holographic Thermalization}},
  \href{http://xxx.lanl.gov/abs/1103.2683}{{\tt arXiv:1103.2683}}. * Temporary
  entry *.

\bibitem{Ebrahim:2010ra}
H.~Ebrahim and M.~Headrick, {\it {Instantaneous Thermalization in Holographic
  Plasmas}},  \href{http://xxx.lanl.gov/abs/1010.5443}{{\tt arXiv:1010.5443}}.

\bibitem{Asplund:2011qj}
C.~Asplund, D.~Berenstein, and D.~Trancanelli, {\it {Evidence for fast
  thermalization in the BMN matrix model}},
  \href{http://xxx.lanl.gov/abs/1104.5469}{{\tt arXiv:1104.5469}}.

\bibitem{Garfinkle:2004pw}
D.~Garfinkle, {\it {Gravitational collapse in anti de Sitter space}},  {\em
  Phys.Rev.} {\bf D70} (2004) 104015,
  [\href{http://xxx.lanl.gov/abs/gr-qc/0408064}{{\tt gr-qc/0408064}}].

\bibitem{Garfinkle:2004sx}
D.~Garfinkle, {\it {Numerical simulation of a possible counterexample to cosmic
  censorship}},  {\em Phys.Rev.} {\bf D69} (2004) 124017,
  [\href{http://xxx.lanl.gov/abs/gr-qc/0403078}{{\tt gr-qc/0403078}}].

\bibitem{Bizon:2011gg}
P.~Bizon and A.~Rostworowski, {\it {On weakly turbulent instability of anti-de
  Sitter space}},  {\em Phys.Rev.Lett.} {\bf 107} (2011) 031102,
  [\href{http://xxx.lanl.gov/abs/1104.3702}{{\tt arXiv:1104.3702}}].

\bibitem{Jalmuzna:2011qw}
J.~Jalmuzna, A.~Rostworowski, and P.~Bizon, {\it {A comment on AdS collapse of
  a scalar field in higher dimensions}},
  \href{http://xxx.lanl.gov/abs/1108.4539}{{\tt arXiv:1108.4539}}. * Temporary
  entry *.

\bibitem{Dias:2011ss}
O.~J. Dias, G.~T. Horowitz, and J.~E. Santos, {\it {Gravitational Turbulent
  Instability of Anti-de Sitter Space}},
  \href{http://xxx.lanl.gov/abs/1109.1825}{{\tt arXiv:1109.1825}}. * Temporary
  entry *.

\bibitem{Danielsson:1999zt}
U.~H. Danielsson, E.~Keski-Vakkuri, and M.~Kruczenski, {\it {Spherically
  collapsing matter in AdS, holography, and shellons}},  {\em Nucl.Phys.} {\bf
  B563} (1999) 279--292, [\href{http://xxx.lanl.gov/abs/hep-th/9905227}{{\tt
  hep-th/9905227}}].

\bibitem{Danielsson:1999fa}
U.~H. Danielsson, E.~Keski-Vakkuri, and M.~Kruczenski, {\it {Black hole
  formation in AdS and thermalization on the boundary}},  {\em JHEP} {\bf 0002}
  (2000) 039, [\href{http://xxx.lanl.gov/abs/hep-th/9912209}{{\tt
  hep-th/9912209}}].

\bibitem{Giddings:1999zu}
S.~B. Giddings and S.~F. Ross, {\it {D3-brane shells to black branes on the
  Coulomb branch}},  {\em Phys.Rev.} {\bf D61} (2000) 024036,
  [\href{http://xxx.lanl.gov/abs/hep-th/9907204}{{\tt hep-th/9907204}}].

\bibitem{Giddings:2001ii}
S.~B. Giddings and A.~Nudelman, {\it {Gravitational collapse and its boundary
  description in AdS}},  {\em JHEP} {\bf 0202} (2002) 003,
  [\href{http://xxx.lanl.gov/abs/hep-th/0112099}{{\tt hep-th/0112099}}].

\bibitem{Alberghi:2003ce}
G.~Alberghi, R.~Casadio, and G.~Venturi, {\it {Thermodynamics for radiating
  shells in anti-de Sitter space-time}},  {\em Phys.Lett.} {\bf B557} (2003)
  7--11, [\href{http://xxx.lanl.gov/abs/gr-qc/0302038}{{\tt gr-qc/0302038}}].

\bibitem{Alberghi:2003pr}
G.~Alberghi and R.~Casadio, {\it {On the gravitational collapse in anti-de
  Sitter space-time}},  {\em Phys.Lett.} {\bf B571} (2003) 245--249,
  [\href{http://xxx.lanl.gov/abs/gr-qc/0306002}{{\tt gr-qc/0306002}}].

\bibitem{Birmingham:1999yt}
D.~Birmingham and S.~Sen, {\it {Gott time machines, BTZ black hole formation,
  and Choptuik scaling}},  {\em Phys.Rev.Lett.} {\bf 84} (2000) 1074--1077,
  [\href{http://xxx.lanl.gov/abs/hep-th/9908150}{{\tt hep-th/9908150}}].

\bibitem{Husain:2000vm}
V.~Husain and M.~Olivier, {\it {Scalar field collapse in three-dimensional AdS
  space-time}},  {\em Class.Quant.Grav.} {\bf 18} (2001) L1--L10,
  [\href{http://xxx.lanl.gov/abs/gr-qc/0008060}{{\tt gr-qc/0008060}}].

\bibitem{Pretorius:2000yu}
F.~Pretorius and M.~W. Choptuik, {\it {Gravitational collapse in
  (2+1)-dimensional AdS space-time}},  {\em Phys.Rev.} {\bf D62} (2000) 124012,
  [\href{http://xxx.lanl.gov/abs/gr-qc/0007008}{{\tt gr-qc/0007008}}].

\bibitem{Birmingham:2001hc}
D.~Birmingham, {\it {Choptuik scaling and quasinormal modes in the AdS / CFT
  correspondence}},  {\em Phys.Rev.} {\bf D64} (2001) 064024,
  [\href{http://xxx.lanl.gov/abs/hep-th/0101194}{{\tt hep-th/0101194}}].

\bibitem{Husain:2002nk}
V.~Husain, G.~Kunstatter, B.~Preston, and M.~Birukou, {\it {Anti-deSitter
  gravitational collapse}},  {\em Class. Quant. Grav.} {\bf 20} (2003)
  L23--L30, [\href{http://xxx.lanl.gov/abs/gr-qc/0210011}{{\tt
  gr-qc/0210011}}].

\bibitem{Birukou:2002ge}
M.~Birukou, V.~Husain, G.~Kunstatter, E.~Vaz, and M.~Olivier, {\it {Spherically
  symmetric scalar field collapse in any dimension}},  {\em Phys. Rev.} {\bf
  D65} (2002) 104036.

\bibitem{Witek:2010zz}
H.~Witek, V.~Cardoso, C.~Herdeiro, A.~Nerozzi, U.~Sperhake, {\em et.~al.}, {\it
  {Black holes in a box}},  {\em J.Phys.Conf.Ser.} {\bf 229} (2010) 012072.

\bibitem{Witek:2010qc}
H.~Witek, V.~Cardoso, C.~Herdeiro, A.~Nerozzi, U.~Sperhake, {\em et.~al.}, {\it
  {Black holes in a box: towards the numerical evolution of black holes in
  AdS}},  {\em Phys.Rev.} {\bf D82} (2010) 104037,
  [\href{http://xxx.lanl.gov/abs/1004.4633}{{\tt arXiv:1004.4633}}].

\bibitem{Chesler:2008hg}
P.~M. Chesler and L.~G. Yaffe, {\it {Horizon formation and far-from-equilibrium
  isotropization in supersymmetric Yang-Mills plasma}},  {\em Phys.Rev.Lett.}
  {\bf 102} (2009) 211601, [\href{http://xxx.lanl.gov/abs/0812.2053}{{\tt
  arXiv:0812.2053}}].

\bibitem{Chesler:2009cy}
P.~M. Chesler and L.~G. Yaffe, {\it {Boost invariant flow, black hole
  formation, and far-from-equilibrium dynamics in N = 4 supersymmetric
  Yang-Mills theory}},  {\em Phys.Rev.} {\bf D82} (2010) 026006,
  [\href{http://xxx.lanl.gov/abs/0906.4426}{{\tt arXiv:0906.4426}}].

\bibitem{Sotiriadis:2010si}
S.~Sotiriadis and J.~Cardy, {\it {Quantum quench in interacting field theory: a
  self- consistent approximation}},  {\em Phys. Rev.} {\bf B81} (2010) 134305,
  [\href{http://xxx.lanl.gov/abs/1002.0167}{{\tt arXiv:1002.0167}}].

\bibitem{Calabrese:2007rg}
P.~Calabrese and J.~Cardy, {\it {Quantum Quenches in Extended Systems}},  {\em
  J. Stat. Mech.} {\bf 0706} (2007) P06008,
  [\href{http://xxx.lanl.gov/abs/0704.1880}{{\tt arXiv:0704.1880}}].

\bibitem{Heller:2011ju}
M.~P. Heller, R.~A. Janik, and P.~Witaszczyk, {\it {The characteristics of
  thermalization of boost-invariant plasma from holography}},
  \href{http://xxx.lanl.gov/abs/1103.3452}{{\tt arXiv:1103.3452}}. * Temporary
  entry *.

\bibitem{Janik:2005zt}
R.~A. Janik and R.~B. Peschanski, {\it {Asymptotic perfect fluid dynamics as a
  consequence of Ads/CFT}},  {\em Phys.Rev.} {\bf D73} (2006) 045013,
  [\href{http://xxx.lanl.gov/abs/hep-th/0512162}{{\tt hep-th/0512162}}].

\bibitem{Bjorken:1982qr}
J.~Bjorken, {\it {Highly Relativistic Nucleus-Nucleus Collisions: The Central
  Rapidity Region}},  {\em Phys.Rev.} {\bf D27} (1983) 140--151.

\bibitem{Chesler:2010bi}
P.~M. Chesler and L.~G. Yaffe, {\it {Holography and colliding gravitational
  shock waves in asymptotically AdS$_5$ spacetime}},  {\em Phys.Rev.Lett.} {\bf
  106} (2011) 021601, [\href{http://xxx.lanl.gov/abs/1011.3562}{{\tt
  arXiv:1011.3562}}].

\bibitem{Balasubramanian:1999re}
V.~Balasubramanian and P.~Kraus, {\it {A Stress tensor for Anti-de Sitter
  gravity}},  {\em Commun.Math.Phys.} {\bf 208} (1999) 413--428,
  [\href{http://xxx.lanl.gov/abs/hep-th/9902121}{{\tt hep-th/9902121}}].

\bibitem{deHaro:2000xn}
S.~de~Haro, S.~N. Solodukhin, and K.~Skenderis, {\it {Holographic
  reconstruction of space-time and renormalization in the AdS / CFT
  correspondence}},  {\em Commun.Math.Phys.} {\bf 217} (2001) 595--622,
  [\href{http://xxx.lanl.gov/abs/hep-th/0002230}{{\tt hep-th/0002230}}].

\bibitem{Back:2004je}
B.~Back, M.~Baker, M.~Ballintijn, D.~Barton, B.~Becker, {\em et.~al.}, {\it
  {The PHOBOS perspective on discoveries at RHIC}},  {\em Nucl.Phys.} {\bf
  A757} (2005) 28--101, [\href{http://xxx.lanl.gov/abs/nucl-ex/0410022}{{\tt
  nucl-ex/0410022}}]. PHOBOS White Paper on discoveries at RHIC.

\bibitem{Aamodt:2010pa}
{\bf The ALICE Collaboration} Collaboration, K.~Aamodt {\em et.~al.}, {\it
  {Elliptic flow of charged particles in Pb-Pb collisions at 2.76 TeV}},  {\em
  Phys.Rev.Lett.} {\bf 105} (2010) 252302,
  [\href{http://xxx.lanl.gov/abs/1011.3914}{{\tt arXiv:1011.3914}}].

\bibitem{Aamodt:2010pb}
{\bf The ALICE Collaboration} Collaboration, B.~Abelev {\em et.~al.}, {\it
  {Charged-particle multiplicity density at mid-rapidity in central Pb-Pb
  collisions at $\sqrt{s_{NN}} = 2.76$ TeV}},  {\em Phys.Rev.Lett.} {\bf 105}
  (2010) 252301, [\href{http://xxx.lanl.gov/abs/1011.3916}{{\tt
  arXiv:1011.3916}}].

\bibitem{Minic:2008xa}
D.~Minic and M.~Pleimling, {\it {Non-relativistic AdS/CFT and Aging/Gravity
  Duality}},  {\em Phys. Rev.} {\bf E78} (2008) 061108,
  [\href{http://xxx.lanl.gov/abs/0807.3665}{{\tt arXiv:0807.3665}}].

\bibitem{Nakayama:2010xq}
Y.~Nakayama, {\it {Universal time-dependent deformations of Schrodinger
  geometry}},  {\em JHEP} {\bf 04} (2010) 102,
  [\href{http://xxx.lanl.gov/abs/1002.0615}{{\tt arXiv:1002.0615}}].

\bibitem{Jottar:2010vp}
J.~I. Jottar, R.~G. Leigh, D.~Minic, and L.~A. Pando~Zayas, {\it {Aging and
  Holography}},  {\em JHEP} {\bf 11} (2010) 034,
  [\href{http://xxx.lanl.gov/abs/1004.3752}{{\tt arXiv:1004.3752}}].

\end{thebibliography}\endgroup

\end{document}